\documentclass[%
 aip,
 jmp,%
 amsmath,amssymb,
 preprint,
reprint,%
 nofootinbib,%
]{revtex4-1}

\makeatletter
\renewcommand\@make@capt@title[2]{%
 \@ifx@empty\float@link{\@firstofone}{\expandafter\href\expandafter{\float@link}}%
  {\textbf{#1}}\@caption@fignum@sep#2\quad
}%
\makeatother
\usepackage{soul,color}
\usepackage{graphicx}
\usepackage{amssymb}
\usepackage{epstopdf, epsfig}
\usepackage{epstopdf}
\AppendGraphicsExtensions{.tif}
\usepackage[rightcaption]{sidecap} 
\usepackage{subfigure}
\usepackage{multirow}
\usepackage{amsmath}
\usepackage{mathptmx}
\usepackage{flafter}
\usepackage{natbib}
\usepackage{textcmds}
\usepackage[colorlinks,citecolor=blue,urlcolor=blue,bookmarks=false,hypertexnames=true]{hyperref}

\begin{document}


\title{Identifying optimal location for control of thermoacoustic instability through statistical analysis of saddle point trajectories}

\author{C.P. Premchand}
\thanks{Author to whom correspondence should be addressed}
\email{premchand.iitb@gmail.com}
\affiliation{Department of Aerospace Engineering, Indian Institute of Technology - Bombay, 400076 - Mumbai, India}
 
\author{Abin Krishnan}
\affiliation{%
Department of Aerospace Engineering, Indian Institute of Technology - Madras, 600036 - Chennai, India
}%

\author{Manikandan Raghunathan}
\affiliation{%
Department of Aerospace Engineering, Indian Institute of Technology - Madras, 600036 - Chennai, India
}%

\author{P.R. Midhun}
\affiliation{%
Department of Aerospace Engineering, Indian Institute of Technology - Madras, 600036 - Chennai, India
}%

\author{Reeja K.V.}
\affiliation{%
Department of Aerospace Engineering, Indian Institute of Technology - Madras, 600036 - Chennai, India
}%

\author{R.I. Sujith}
\affiliation{%
Department of Aerospace Engineering, Indian Institute of Technology - Madras, 600036 - Chennai, India
}%

\author{Vineeth Nair}
\thanks{Author to whom correspondence should be addressed}
\email{vineeth@aero.iitb.ac.in}
\affiliation{%
Department of Aerospace Engineering, Indian Institute of Technology - Bombay, 400076 - Mumbai, India
}%

\begin{abstract}
We propose a framework of Lagrangian Coherent Structures (LCS) to enable passive open-loop control of tonal sound generated during thermoacoustic instability. Experiments were performed in a laboratory-scale bluff-body stabilized turbulent combustor in the state of thermoacoustic instability. We use dynamic mode decomposition (DMD) on the flow-field to identify dynamical regions where the acoustic frequency is dominant. We find that the separating shear layer from the backward-facing step of the combustor envelops a cylindrical vortex in the outer recirculation zone (ORZ), which eventually impinging on the top wall of the combustor during thermoacoustic instability. 
We track the saddle points in this shear layer emerging from the backward facing step over several acoustic cycles. A passive control strategy is then developed by injecting a steady stream of secondary air targeting the identified optimal location where the saddle points spend a majority of their time in a statistical sense.

After implementing the control action, the resultant flow-field is also analysed using LCS to understand the key differences in flow dynamics. We find that the shear layer emerging from the dump plane is deflected in a direction almost parallel to the axis of the combustor after the control action. This deflection in turn prevents the shear layer from enveloping the vortex and impinging on the combustor walls, resulting in a drastic reduction in the amplitude of the sound produced.
\end{abstract}


\maketitle

\begin{quotation}
The increasing worldwide adoption of natural gas for the purpose of power generation has intensified the need for cleaner and more efficient combustion processes. However, this transition towards lean combustion, while fuel-efficient, has given rise to a daunting challenge: thermoacoustic instability. The occurrence of this phenomenon results in the emergence of high-amplitude pressure oscillations within combustion systems, often leading to catastrophic failures in engines and electronics. The complex interaction between flame oscillations and acoustic modes in combustors has perplexed researchers and engineers, presenting challenges in finding direct solutions due to its system-dependent nature. In order to address this significant issue, researchers have explored various active and passive control strategies, seeking to disrupt the flame-acoustic coupling responsible for thermoacoustic instability. One promising avenue is the identification of optimal locations within the combustor to introduce secondary micro-jets that can mitigate instability. While previous research has delved into data-driven approaches to locate these optimal points, this novel study takes a different path. By delving deep into the flow dynamics within the thermoacoustic system, we aim to pinpoint optimal locations through the framework of Lagrangian Coherent Structures (LCS). This groundbreaking patented approach promises to provide a more comprehensive understanding of the complex flow physics governing thermoacoustic instability, potentially unlocking effective control solutions.
\end{quotation}

\section{Introduction}
\label{INTRODUCTION}
In recent years, natural gas (23.5\% of the global share of electricity production in 2019\footnote{https://www.iea.org/data-and-statistics/charts/world-gross-electricity-production-by-source-2019}) is increasingly being used to generate electricity all around the globe. Due to environmental concerns of climate change, industrial gas turbine engine manufacturers are pushed to develop low emission combustors which are now a reliable source for evergrowing demands in electricity. Although lean combustion is fuel efficient, it creates serious issues such as thermoacoustic instability and flame destabilization leading to blow-out. 

Thermoacoustic instability manifests as large amplitude periodic pressure oscillations. These high-amplitude periodic pressure oscillations are a resultant of positive feedback between the oscillations of the flame with one of the acoustic modes of the combustor \citep{rayleigh}. The oscillations, manifested as tonal sound, are self-sustained, often leading to a sudden failure of engine components and electronics \citep{Juniper}. Many rockets \citep{Sutton} and gas turbine engines \citep{Lieuwen2} have been catastrophically destroyed due to thermoacoustic instability. In fact, thermoacoustic instability was one of the major challenges faced in the development of rocket engines for the Apollo moon mission \citep{Biggs}. Although researchers and engineers in the field have come up with ad-hoc solutions to mitigate instability, the root causes are not well understood as they are often system dependent and not amenable to generalization.

The turbulent bluff-body stabilized dump combustor presented in this study exhibits thermoacoustic instability in lean combustion conditions. There is a transition of dynamics from a stable state of combustion noise to a state of unstable operation (thermoacoustic instability) via an intermittency route \citep{Nair5, nair}. Combustion noise is an operational state that has a broadband signature in the state variable of the system\citep{Candel}. Thermoacoustic instability, on the other hand, is an operational state that exhibits narrowband signatures. The transitional state\textemdash intermittency\textemdash shows both signatures of combustion noise and thermoacoustic instability. Other studies on this combustor are summarized by Sujith and Unni\citep{Sujith_PoF, Sujith_proceedings} in recent review papers. Such intermittent routes to instability have also been observed in combustors with other types of flame-holding mechanisms \citep{gotoda, Ebi, kheirkhah, sampath_CaF}. Coherent structures play a vital role in the production of the sound during the transition to thermoacoustic instability.

Researchers have attempted various ways of disrupting the flame\textendash acoustic coupling to control thermoacoustic instability, both actively \citep{Zhao1} and passively \citep{Zhao2}. Active control measures involve modulation of the air/fuel flow rates through actuators, tunable valves and loudspeakers \citep{McManus}. Though active control measures are successful across a wide range of operating conditions, use of electro-mechanical components adds complexity for practical implementation. Unlike active control, passive control uses a slight modification to the geometry of the combustor; for example, modification of the geometry, location of the air or fuel injector \citep{Ghoniem, Oztarlik}, introduction of baffles, steady micro-jet injection or by using damping devices such as Helmholtz resonators or acoustic liners \citep{Schadow1, Altay_CaF}. Though passive control measures are simple and robust, they are mostly based on a trial-and-error approach. Thus, controlling thermoacoustic instability is still a challenging problem in gas turbines and rocket engines.

In most passive control strategies such as secondary micro-jet injection, the location to implement the control strategy is of prime importance. Identifying optimal locations in the combustor would drastically improve the control effectiveness. The region in the flow-field that dictates the spatio-temporal dynamics during thermoacoustic instability is one such optimal location\cite{poinsot}. A secondary microjet of air/fuel can be introduced in such optimal locations to disrupt the spatio-temporal patterns and mitigate instability. 

In previous studies in the literature, Rayleigh index distribution \citep{Tachibana} and flame anchoring regions \citep{Ghoniem, Altay_CaF} were considered as optimal locations for injecting secondary fuel. There are also other approaches to find the optimal locations employing complex network \citep{Unni_chaos, Abin, Abin1, Abin2, Tandon, Sujith_PoF, Sujith_proceedings} and Hurst exponent measures \citep{Roy_CaF2}. The main emphasis of these papers was on utilising data-driven approaches to examine thermoacoustic systems. In contrast, our primary focus lies in analyzing the flow dynamics within the thermoacoustic system. These techniques did not focus on flow physics; instead, they focused on the emergence of correlations and periodicity in an underlying chaotic flow-field. It is hard to characterise or quantify the flow physics of the system because of the highly complex spatio-temporal dynamics. Therefore, we utilize the framework of Lagrangian Coherent Structures (LCS) along with a statistical analysis to identify the optimal locations on tracked fluid trajectories to achieve control of instability.

According to Haller\citep{Haller}, Lagrangian coherent structures or LCS are locally the most attracting or repelling material lines or surfaces in the flow-field. For a heuristic determination of LCS, the location of fluid parcels are tracked in time to generate a flow map. Then, extracting the ridges in the maximum eigenvalue field (called the finite-time Lyapunov exponent (FTLE) field) of the Jacobian of this flow map, one obtains the repelling LCS. The attracting LCS are identified as the ridges in the FTLE field when the fluid parcels are evolved backwards in time. Although the ridges of FTLE field are only an approximation to LCS, they give a fairly good visual indication of the instantaneous location and shape of the corresponding material lines or surfaces for most practical applications. 

Recently, it was shown that exact LCS computations are possible from the eigenvector field of the maximum eigenvalue using the approach outlined in Farazmand and Haller \citep{Farazmand}. The LCS thus identified when the fluid parcels are evolved forwards or backwards in time are correspondingly referred to as repelling LCS and attracting LCS. The first major objective of our study is to present this accurate LCS framework for the first time in experimental data obtained from a moderately complex, turbulent, reacting flow-field inside a laboratory-scale combustor.

It was previously reported using the LCS framework with FTLE fields that periodic oscillations in the shear layers emerging from the dump plane and from the top of the bluff-body are responsible for thermoacoustic instability for the combustor studied in this paper \citep{prem_pof}. The superiority of the LCS framework was established by presenting a complete visualization of shear layers as they evolve in time which is not possible with other vortex tracking methods. In this paper, we show that an optimal location along the shear layer for passive control can be obtained if we identify the trajectory of saddle points. The system and method to identify optimal locations using saddle points for control action have been patented\cite{prem_patent}. In a recent study, saddle-type flows were identified using LCS framework with FTLE fields that separates the dynamics between the different combustor lengths\cite{Sampath}. saddle points are identified as the common points when repelling and attracting are overlaid on top of each other at a given time. These saddle points dictate the flow and hence flame dynamics at any given time and as they advect, they create the two shear layers responsible for sound production during thermoacoustic instability\cite{prem_asme}. Tracking of trajectories of saddle points followed by a statistical analysis allow us to identify the optimal locations along the shear layer for passive control. The identification of the optimal locations in the combustor then forms the second major objective of our study.

Validation of the methodology will be performed by injecting a secondary micro-jet of air at the estimated optimal location to disrupt the observed trajectory of the identified saddle points. We will then compare and contrast the saddle point trajectories and the flow dynamics before and after control action. This then forms the third and final major objective of our study.

The paper is organised as follows: the experimental setup and data acquisition carried out in the study are explained in Section \ref{EXPERIMENTAL SETUP AND DATA ACQUISITION}. The method to accurately compute Lagrangian Coherent Structures is elucidated in Section \ref{METHODOLOGY}. We discuss the spatio-temporal flow features observed during the state of thermoacoustic instability and the suppressed state after control action in Section \ref{RESULTS AND DISCUSSION}. We summarize the key results in Section \ref{CONCLUSIONS}. 

\section{Experimental setup and data acquisition}
\label{EXPERIMENTAL SETUP AND DATA ACQUISITION}
\begin{figure*}
\centering
\includegraphics[scale=0.12]{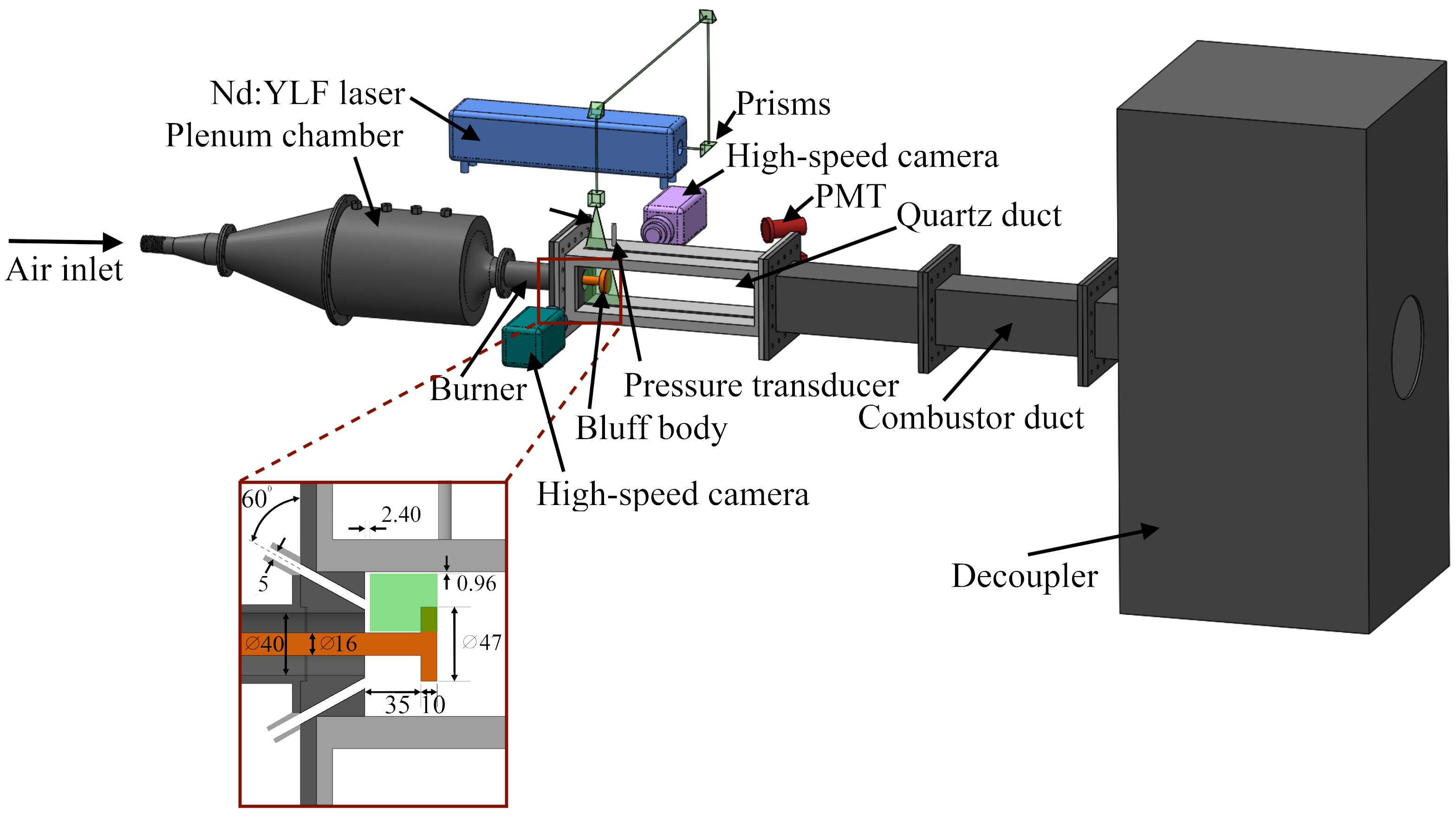}
\caption{Schematic of the experimental setup used in the study. The geometry of the bluff-body and the burner upstream are shown in the inset. Secondary micro-jet ports are placed at an inclination of $60{^\circ}$ from the vertical of the dump plane. The green-colored box represents the region where spatial data is acquired.}
\label{fig:Exp_setup_passive_control_modified_initial}
\end{figure*}

Experiments were performed in a dump combustor with a bluff-body to stabilize the turbulent flame (see figure \ref{fig:Exp_setup_passive_control_modified_initial}). The bluff-body is a circular disk with a diameter of 47 mm and thickness 10 mm, located 35 mm from the dump plane, mounted on a shaft of diameter 16 mm. Four radial injection holes of 1.7 mm are located 120 mm upstream of the dump plane that delivers the fuel into the burner section. The fuel used is Liquified petroleum gas (LPG) with a composition of 60\% butane and 40\% propane by volume. A spark plug connected to an 11 kV transformer mounted on the dump plane ignites the partially premixed fuel-air mixture. The combustor has a cross-section of $90 \times 90$ mm$^{2}$ with an optical window (400 $\times$ 90 mm$^{2}$) for flame imaging and flow visualization. A decoupler of size 1000 $\times$ 500 $\times$ 500 mm$^{3}$ is attached to the end of the combustion chamber (of length 1100 mm) to minimize acoustic radiation losses. Open acoustic boundary condition ($p'=0$) is approximately achieved at the end of the combustion chamber due to the decoupler.

The flow rates of air and fuel are controlled using mass flow controllers (Alicat Scientific, MCR Series) with an uncertainty of $\pm$ 0.8\% of the reading and 0.2\% of full scale. This results in a maximum experimental uncertainty in equivalence ratio, velocity, and Reynolds number of $\pm$ 0.02, $\pm$ 0.01 m/s, and $\pm$ 0.01 respectively. Unsteady pressure fluctuations are measured using a piezoelectric transducer (PCB 103B02 with an uncertainty of $\pm$ 0.15 Pa) mounted 20 mm from the dump plane on the wall of the combustor. The transducer is not flush mounted; as a result, there is a small acoustic phase delay of 5$^\circ$, which is not large enough to affect the analysis and the results presented in this study. We determined the phase shift of $\pm$5$^\circ$ by measuring the phase difference between the calibrated pressure sensor and a reference sensor. This comparison was made while the calibrated sensor was mounted with a T-joint, covering a broad spectrum of frequencies.

The local heat release rate fluctuations are captured using a high-speed CMOS camera (Phantom - v 12.1) and the global heat release rate fluctuations are captured using a photomultiplier tube (PMT) both fitted with a bandpass filter centered at around 435 nm (10 nm FWHM) to capture CH* chemiluminescence \citep{hardalupas}. The high-speed camera is outfitted with a ZEISS 50 mm lens having an aperture of $f/5.6$ that focuses on a region spanning $87$ mm $\times 78$ mm ($800 \times 720$ px) close to the bluff-body. 

Velocity measurements are obtained during the experiments by performing high-speed two-dimensional, two-component particle image velocimetry (2D-2C PIV) using an Nd-YLF laser (single cavity-twin pulsed laser from Photonics) with an operating wavelength of 527 nm. TiO$_{2}$ particles (Kronos make, product - 1071) with an approximate size of 1 $\mu$m are seeded in the reacting flow-field.

More details regarding the experimental setup and data acquisition can be found in Krishnan ${et.al}$\citep{Abin1} and George ${et.al}$\citep{nitin}. Further discussion on the spatiotemporal dynamics of bluff-body stabilized combustors can be found in these articles\citep{nitin, poinsot, prem_pof, chakravarthy, prem_asme}.

\section{Methodology}
\label{METHODOLOGY}
According to the variational LCS theory developed by Haller \citep{Haller}, Lagrangian coherent structures (LCS) are material lines that act as organising patterns in observed trajectories. They distinguish themselves by repelling or attracting the neighbouring trajectories at locally the highest rate in the flow-field. Repelling LCS refers to the material lines across which the fluid parcels diverge, whereas attracting LCS are the material lines along which the fluid parcels converge. 

To compute LCS from a 2D velocity flow-field, the displacement of the fluid parcels are tracked forward in time with reference to an initial position $ \textbf{x}_{0}$ and time $t_{0}$ as $F_{t_{0}}^{t} ( \textbf{x}_{0}):=x(t;\,t_{0},\, \textbf{x}_{0})$, also referred to as a flow map (see figure \ref{fig:Parcel_backward_a_b}). The gradient of the flow map is then obtained by computing the Jacobian $\nabla F_{t_{0}}^{t} ( \textbf{x}_{0})$. 
\begin{figure*}
\centering
\includegraphics[scale=0.5]{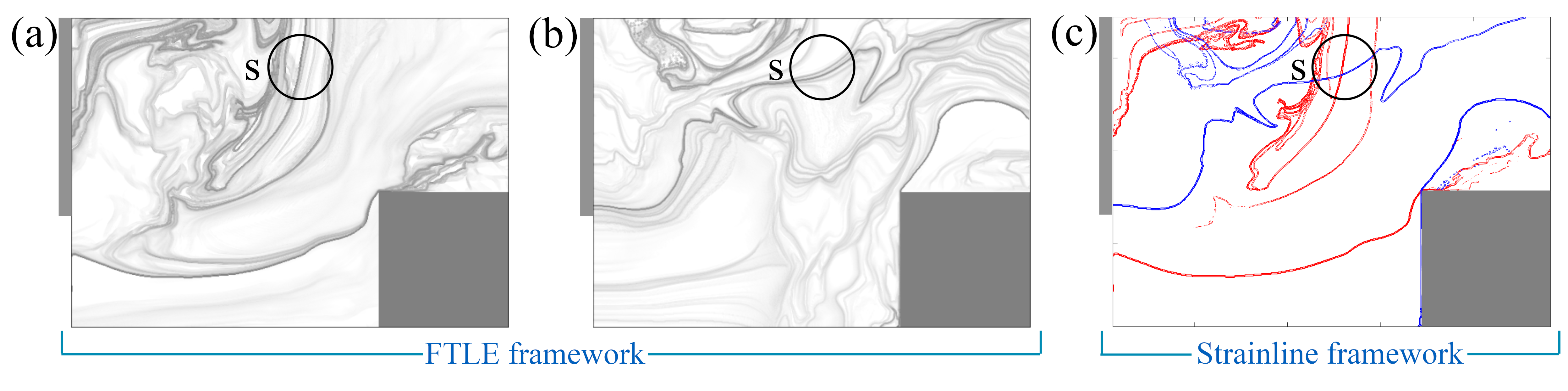}
\caption{(a) Ridges of FTLE field obtained by fluid parcel advection in forward-time showing the repelling LCS; (b) Ridges of FTLE field obtained by fluid parcel advection in backward-time showing the attracting LCS; (c) Attracting LCS (blue) and repelling LCS (red) obtained using strainline framework. Saddle point is marked using a black circle and dump plane is represented by a thick line on the vertical axis of each snapshot. The flow is from left to right. The bluff-body is shaded in grey.}
\label{fig:Strainlines_Stretchlines_rep_rate_MJ0_forward_backward_2070_saddle_labels_modified}
\end{figure*}
\begin{figure*}
\centering
\includegraphics[scale=1.2]{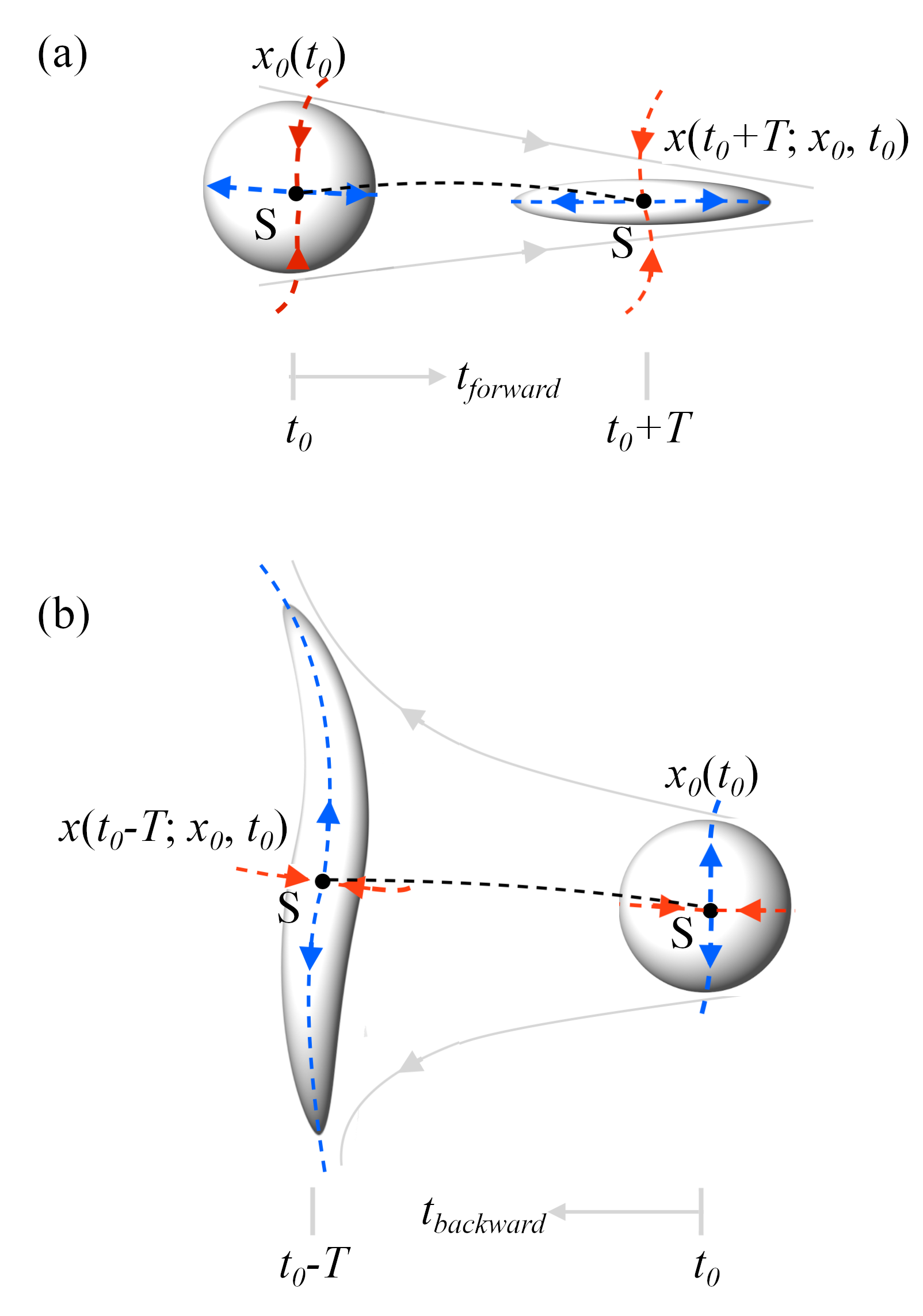}
\caption{A cartoon depicting the evolution of a fluid parcel containing the saddle point (marked with a black dot and labelled “S’) in a Eulerian frame of reference. The saddle point trajectory is shown using a black dashed line. The dashed blue and red lines are attracting LCS and repelling LCS, respectively. (a) Neighbouring fluid trajectories on either side of an attracting LCS constructed from the $\rho_{t_{0}}^{t_{0}-T}$ field attract when advected forward in time; i.e., a fluid parcel with initial position $ \textbf{x}_{0}$ and time $t_{0}$ shrinks upon advecting in forward in time. (b) Neighbouring fluid parcels on either side of a repelling LCS attract when advected forward in time or repel when advected backward in time; i.e., a fluid parcel with initial position $ \textbf{x}_{0}$ and time $t_{0}$ stretches as it advects backward in time.}
\label{fig:Parcel_backward_a_b}
\end{figure*}


The framework outlined in Farazmand and Haller\citep{Farazmand} is utilized to extract exact LCS. Consider a material line $\mathcal{M}$ that evolves in time as $\mathcal{M}(t)= F_{t_{0}}^{t} (\mathcal{M}(t_{0}))$. The normal repulsion rate $(\rho_{t_{0}}^{t_{0}+T}(\textbf{x}_{0},\textbf{n}_{0}))$ of the material line is a function of the length of the surface-normal component ($\textbf{n}_{0}$) of the advected vector which can be mathematically formulated as,
\vspace{-1em}
\begin{equation}\label{eq:3}
\rho_{t_{0}}^{t_{0}+T}(\textbf{x}_{0},\textbf{n}_{0})=\frac{1}{\sqrt{\Big<\textbf{n}_{0},\Big[C_{t_{0}}^{t} ( \textbf{x}_{0})\Big]^{-1} \textbf{n}_{0}\Big>}}
\end{equation}
\noindent where the right Cauchy-Green strain tensor $C_{t_{0}}^{t} ( \textbf{x}_{0})$ is defined as,

\vspace{-1em}
\begin{center}
\begin{equation}\label{eq:1}
C_{t_{0}}^{t} ( \textbf{x}_{0},\,t_{0}) = \Big[\nabla F_{t_{0}}^{t} ( \textbf{x}_{0})\Big]^\dagger \nabla F_{t_{0}}^{t} ( \textbf{x}_{0})
\end{equation}
\end{center}
The eigenvalues ($\boldsymbol{\lambda}_{max}$, $\boldsymbol{\lambda}_{min}$) and the eigenvectors ($ \boldsymbol{\xi}_{max}, \boldsymbol{\xi}_{min}$) of $C_{t_{0}}^{t}$ contain information about the most attracting and repelling material lines. The inverse of the square root of the angle between the normalized maximum eigenvector and the component of $C_{t_{0}}^{t} ( \textbf{x}_{0})$ in the direction of $ \boldsymbol{\xi}_{max}$ gives the repulsion rate of the material line at a given location. The strainline framework allows more precise definition of LCS even for shear dominated flows. Further, Jacobian of the flowmap tends to be error prone in shear dominated regions as they don't discriminate stretching from shear. In addition, they are not frame invariant. To circumvent this difficulty, we use the Cauchy-Green strain tensor\citep{Farazmand, Haller}.

To ensure the material lines obtained are LCS, the following mathematical conditions (i-iv) are applied at the grid points on the field of $\rho_{t_{0}}^{t_{0}+T}( \textbf{x}_{0},\textbf{n}_{0})$.

\noindent (i) $\lambda_{min}( \textbf{x}_{0})\neq\lambda_{max}( \textbf{x}_{0})>1;$\\
(ii) $\langle\,  \boldsymbol{\xi}_{max}( \textbf{x}_{0}), \nabla^{2}\lambda_{max}( \textbf{x}_{0}) \boldsymbol{\xi}_{max}( \textbf{x}_{0})\rangle<0;$\\
(iii) $ \boldsymbol{\xi}_{max}( \textbf{x}_{0}) \perp \mathcal{M}(t_{0});$\\
(iv) $\langle\, \nabla\lambda_{max}( \textbf{x}_{0}), \boldsymbol{\xi}_{max}( \textbf{x}_{0})\rangle\ =0;$\\

Applying condition (i) onto the field of maximum repulsion rate, we ensure that the normal repulsion rate $(\rho_{t_{0}}^{t_{0}+T}(\textbf{x}_{0},\textbf{n}_{0}))$) is larger than tangential stretch due to shear along the LCS. Conditions (iii) and (iv) together guarantee that normal repulsion rate $(\rho_{t_{0}}^{t_{0}+T}(\textbf{x}_{0},\textbf{n}_{0}))$) attains a local extremum along the LCS relative to all neighbouring material lines. Finally, condition (ii) ensures that this extremum is a strict local maximum. The resulting contour lines of $\rho_{t_{0}}^{t_{0}+T}(\textbf{x}_{0},\textbf{n}_{0})$ reveal repelling LCS (red lines in figure \ref{fig:Strainlines_Stretchlines_rep_rate_MJ0_forward_backward_2070_saddle_labels_modified}c). The attracting LCS (extracted from $\rho_{t_{0}}^{t_{0}-T}$ field) are estimated by performing the particle advection backwards in time (blue lines in figure \ref{fig:Strainlines_Stretchlines_rep_rate_MJ0_forward_backward_2070_saddle_labels_modified}c). Repelling LCS extracted from $\rho_{t_{0}}^{t_{0}+T}$ field (red lines) and ridges of FTLE fields (repelling LCS) computed forward in time look comparable over a finite-time interval if the ridges are chosen carefully (refer to Figs.\ref{fig:Strainlines_Stretchlines_rep_rate_MJ0_forward_backward_2070_saddle_labels_modified}c \& \ref{fig:Strainlines_Stretchlines_rep_rate_MJ0_forward_backward_2070_saddle_labels_modified}a respectively). However, over a longer time interval, strainlines and stretchlines from normal repulsion rate field with the four mathematical conditions are the best representation of the most repelling and attracting material lines than ridges of FTLE fields \citep{Farazmand}. Further, the LCS from strainline computation directly gets the information from the eigenvector flow-field and does not involve a subjective ridge estimation as was done in the FTLE framework. Hence, the framework of strainlines yields exact LCS.

It is to be noted that the bluff-body stabilized combustor studied is chosen as the flow-field is highly 2D in nature with the out-of-plane component of velocity being much smaller than the two components measured. The algorithm of strainline framework can be used in 3D flows as well. Currently, we are limited by experimental constraints restricting us to this choice of combustor. For analysis, we assume that the flow field is 2D in nature and therefore, the pathlines for FTLE and strainline framework advect on a planar surface. Due to the complex nature of the turbulent flow-field, the attracting and repelling LCS criss-cross each other over short distances in the shear layer resulting in multiple clustered points of intersection at the expected location. saddle points marked using black circles in Figs.\ref{fig:Strainlines_Stretchlines_rep_rate_MJ0_forward_backward_2070_saddle_labels_modified}a-\ref{fig:Strainlines_Stretchlines_rep_rate_MJ0_forward_backward_2070_saddle_labels_modified}c represent the location of this cluster of intersections and will be used as the representative location of the saddle point that dictates the shear layer dynamics in this study. To aid in visualization and interpretation of results, we will only be presenting the attracting LCS along with the saddle points for the remainder of the study.

\vspace{-10pt}
\section{Results and discussions}
\label{RESULTS AND DISCUSSION}

Experiments were performed at an equivalence ratio of $\phi=0.63$ with an average axial velocity $v_{0}$ of 12.3 m/s under which condition the combustor exhibits thermoacoustic instability. Simultaneous unsteady measurements of pressure and global heat release rate fluctuations (using PMT) are acquired at sampling rates of 20 kHz. The spatial velocity flow-field and heat release fluctuations obtained (using high-speed camera capturing CH* chemiluminescence) on the other hand have comparatively lower sampling rates of 2 kHz and 4 kHz respectively.   
\begin{figure*}
\centering
\includegraphics[scale=0.24]{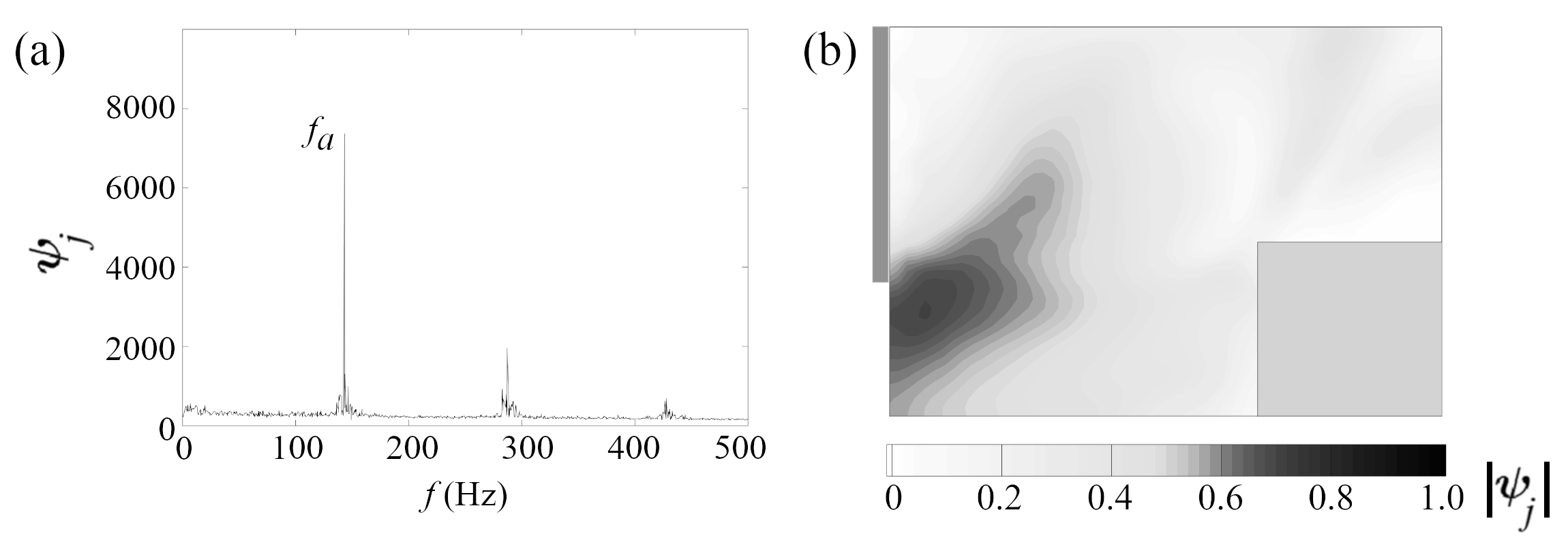}
\caption{(a) Variation of the norm of dynamic modes with frequency. There is a sharp peak at acoustic frequency ($f_{a}$). (b) Dynamic mode at $f_{a}$ showing a region of dominant dynamics upstream of the bluff-body near the dump plane (thick line on the vertical axis). The bluff-body is shaded grey and the flow is from left to right.}
\label{fig:DMNorm_Freq_DM_MJ0}
\end{figure*}
\subsection{Flow dynamics in the state of thermoacoustic instability}
\label{Flow dynamics in the state of thermoacoustic instability}
The velocity field acquired during the state of thermoacoustic instability ($\phi$ = 0.63, $v_{0}$ = 12.3 m/s, $T_0$ = 30 ms) contains dynamics at a well-defined frequency. We perform dynamic mode decomposition (DMD) \citep{Schmid, Schmid1, prem_pof} to extract the spatial mode shape at this frequency. Figure \ref{fig:DMNorm_Freq_DM_MJ0}a shows the variation of the norm of the dynamic modes with frequency during thermoacoustic instability. The dominant frequency is found to be at 143.1 Hz. This dominant frequency corresponds to the tonal sound audible during the state of thermoacoustic instability. Hereafter, we will refer to this frequency as the acoustic frequency ($f_{a}$).

The dynamic mode corresponding to the acoustic frequency ($f_{a}$) is shown in figure \ref{fig:DMNorm_Freq_DM_MJ0}b. This mode is observed to be dominant in the region close to the dump plane where the flow enters the combustor. The existence of such a region implies that there is potentially a sound source responsible for the periodic dynamics in the flow-field generating intense acoustic oscillations during instability. 
\begin{figure*}
    \centering
    \includegraphics[scale=2.12]{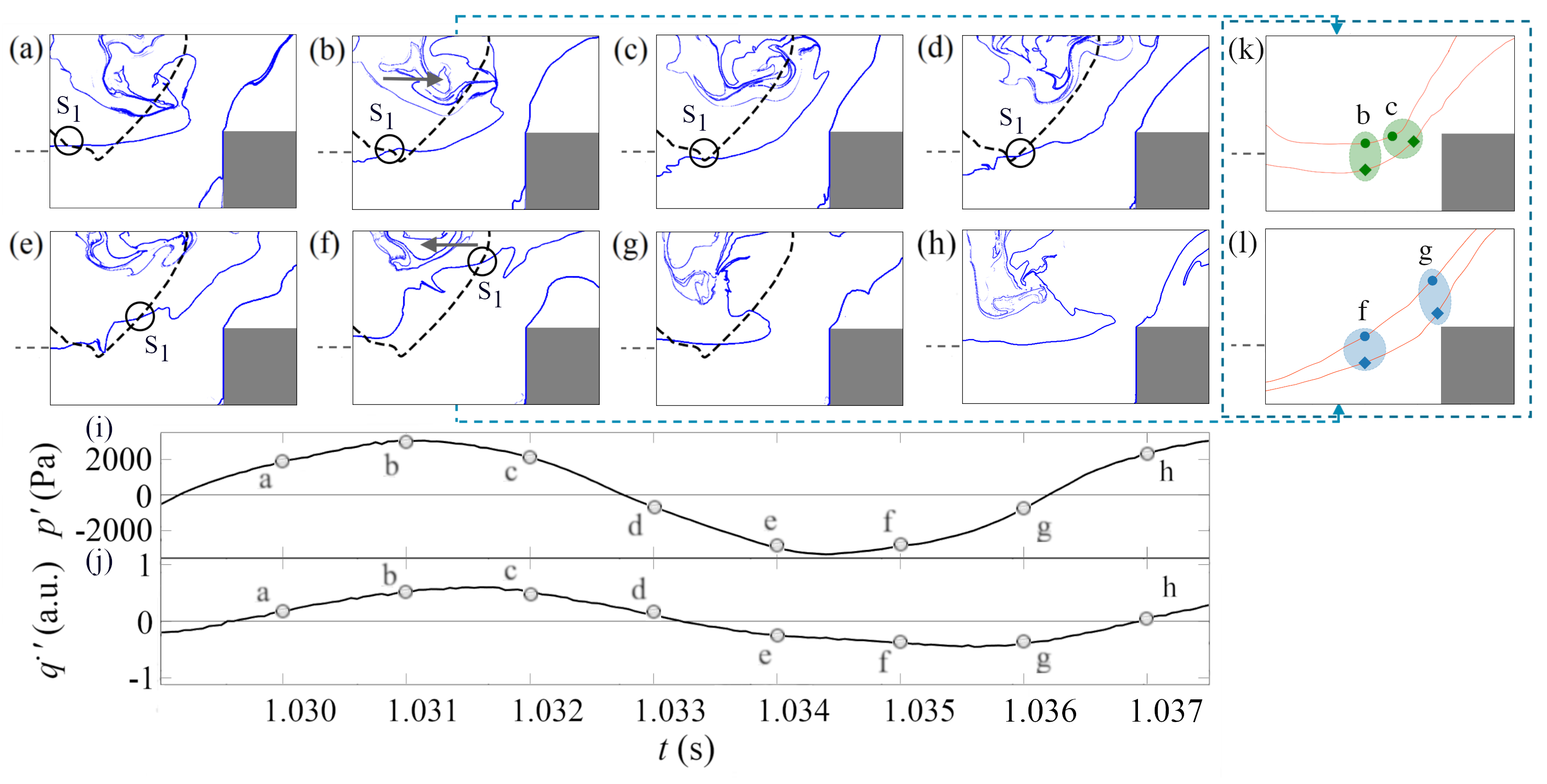}
    \caption{(a-h) Attracting LCS or coherent structures (blue) computed from the velocity data over an acoustic cycle during thermoacoustic instability ($\phi$ = 0.63, $v_{0}$ = 12.3 m/s, $T_0$ = 30 ms) and (i) unsteady pressure and (j) unsteady heat release rate fluctuation signals corresponding to the time instants of the snapshots. The direction in which the vortex moves with respect to the dump plane is shown by a black arrow. The saddle points are marked using black circles in the snapshots. The trajectory of a sample saddle point $S_{1}$ is overlaid using a dashed black line. The pathlines of two fluid parcels during the (k) maximum and (l) minimum of the acoustic cycle. The dump plane is represented by the horizontal dashed line on the vertical axis of each snapshot. The flow is from left to right.}
    \label{fig:Chemi_Instability_wochemi_grayscale_SLPM750_MJ0_without_saddle_circle_labels}
\end{figure*}
We further explore the flow-field using the LCS framework to understand the mechanism of sound production close to the dump plane. The snapshots of attracting LCS (blue lines) computed over one cycle of acoustic oscillation ($T_a=1/f_a$) are presented in Figs. \ref{fig:Chemi_Instability_wochemi_grayscale_SLPM750_MJ0_without_saddle_circle_labels}a-h during the state of thermoacoustic instability. The corresponding time instants are also marked in the time traces of pressure and global heat release rate fluctuations (Figs. \ref{fig:Chemi_Instability_wochemi_grayscale_SLPM750_MJ0_without_saddle_circle_labels}i,j). The attracting LCS computed from $\rho_{t_{0}}^{t_{0}-T}$ field elucidate the shear layer emerging from the dump plane and the vortex bound by this shear layer in the outer recirculation zone (ORZ). The shear layer emerging from the leading edge of the bluff-body is also visible in the snapshots. We focus our attention primarily on the shear layer emerging from the dump plane as this is the region of interest identified by DMD analysis. The corresponding saddle point of interest is identified as the common point between the repelling and attracting LCS emerging from the dump plane (marked with black circles in Fig. \ref{fig:Chemi_Instability_wochemi_grayscale_SLPM750_MJ0_without_saddle_circle_labels} and labelled $S_{1}$). 

We track the evolution of $S_{1}$ over an acoustic cycle in each snapshot. We also separately compute the trajectory of $S_1$ by starting the trajectory at a known location of $S_1$ and evolving forward and backward in time using the known velocity fields (black dashed lines). Since the saddle points must necessarily lie on both the repelling and attracting LCS, the black circles in each snapshot lie along this trajectory. 

The flow-field during thermoacoustic instability consists of three salient features: vortex in the outer recirculation zone (ORZ) and two shear layers, one emerging from the dump plane and another emerging from the top of the bluff-body. The dump plane shear layer envelops the vortex in the ORZ. When the acoustic pressure is rising in the combustor (compression phase, Figs. \ref{fig:Chemi_Instability_wochemi_grayscale_SLPM750_MJ0_without_saddle_circle_labels}e-h) the trapped vortex in ORZ is pushed towards the dump plane as the acoustic velocity is inwards. This motion of vortex moves the shear layer emerging from the dump plane away from the bluff-body. When the acoustic pressure is falling (expansion phase, Figs. \ref{fig:Chemi_Instability_wochemi_grayscale_SLPM750_MJ0_without_saddle_circle_labels}b-d) the vortex in the ORZ is pushed away from the dump plane towards the bluff-body as the acoustic velocity is outwards. This movement of the vortex moves the bounding shear layer towards the bluff-body. The low pressure region is created near the dump plane making the vortex move towards the dump plane during the expansion phase and vice versa.

We track two neighbouring fluid parcels during the maximum (figure \ref{fig:Chemi_Instability_wochemi_grayscale_SLPM750_MJ0_without_saddle_circle_labels} k) and minimum (figure \ref{fig:Chemi_Instability_wochemi_grayscale_SLPM750_MJ0_without_saddle_circle_labels} l) phases of $p'$ respectively. When advected, we see that fluid parcel trajectories approach each other when $p\prime$ is at a maximum and move away from each other when $p\prime$ is at a minimum. Therefore, we conclude that when $p\prime$ is at its minima, acoustic velocity is in the direction of mean flow accelerating the fluid parcels and making them approach each other. When $p\prime$ is at its maxima, acoustic velocity is opposite to the direction of mean flow velocity slowing down the fluid parcels and increasing their separation.

In a previous study, we have reported for the same combustor that the mechanism of tonal sound production during thermoacoustic instability and bursting stages of intermittency is due to the squeezing of the fluid in the region between the two shear layers, one emerging from the dump plane and one from the leading edge of the bluff-body, in a periodic manner, thus creating a monopole sound source\citep{prem_pof}. The above analysis illustrates that the periodic shear layer dynamics is influenced by the movement of the trapped vortex in the ORZ which also happens in phase with the pressure fluctuations inside the combustor. This movement in turn imparts periodicity also to the shear layer emerging from the leading edge of the bluff-body.

Disrupting the periodic behaviour in either one of these shear layers should lead to the suppression of tonal sound. In this study, we focus only on the free shear layer that emerges from the dump plane in the upstream portion of the bluff-body; which was also highlighted through DMD analysis. We use the trajectories containing saddle points that move along the shear layer to identify the \qq{optimal location} to perform passive control. 

\begin{figure*}
        \centering
        \includegraphics[scale=0.402]{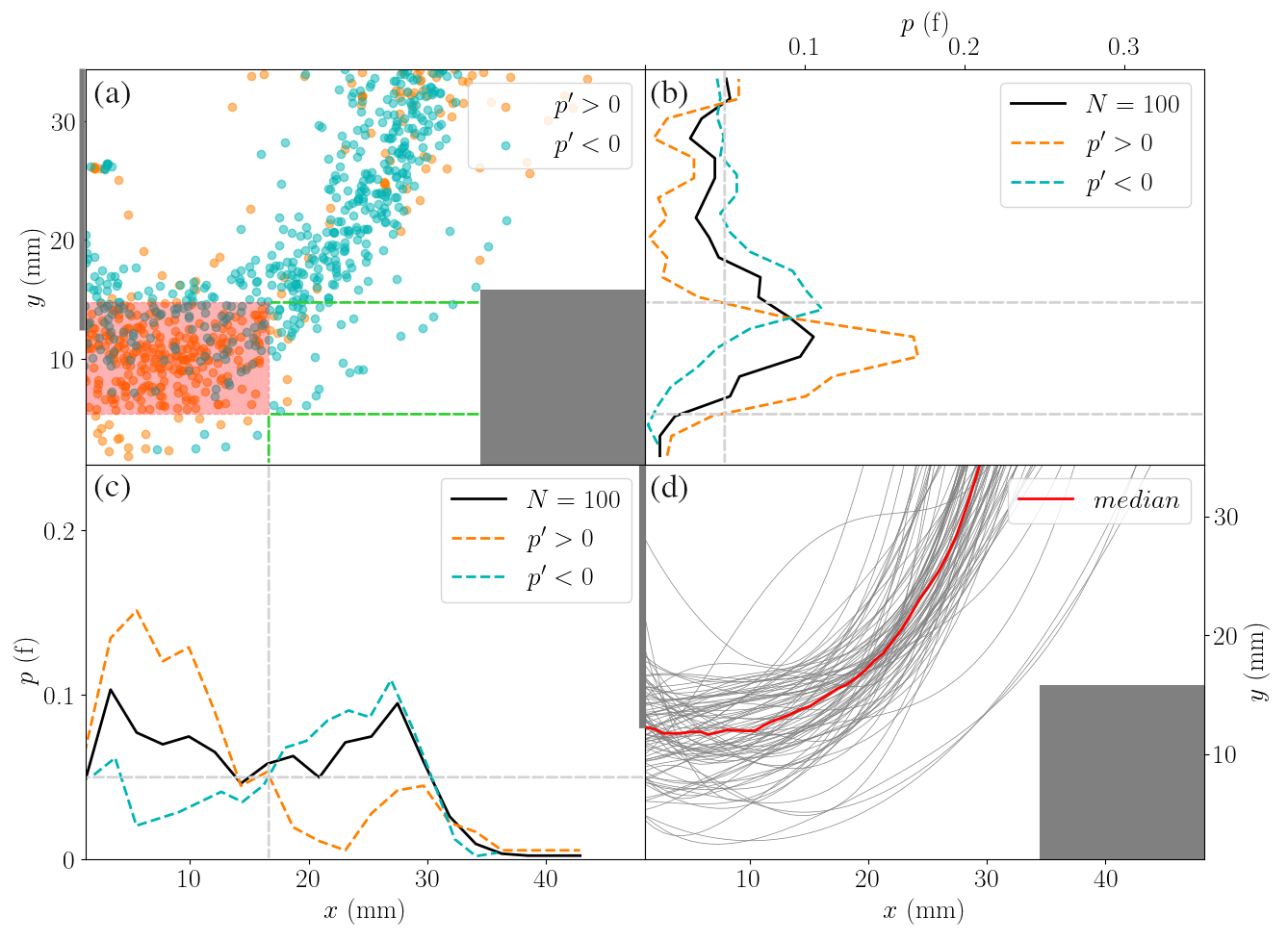}
        \caption{(a) Saddle point locations along the shear layer color-coded based on the sign of pressure fluctuations. Most of the saddle points lie within the red-colored box as the saddle points hardly move during the compression phase of pressure fluctuations. The probability density function (pdf) of the location of saddle points along the (b) vertical direction and (c) horizontal direction provides the location where most of the saddle points are clustered. The pdf of saddle points during $p^\prime>0$ along the vertical and horizontal axis with a threshold bound sets the dimension for the red-boxed region. The bounded lines are extended as gray dashed lines. (d) The trajectories ($N$ = 100) of individual saddle points along the shear layer during thermoacoustic instability. The curvilinear median of these individual trajectories is also shown in red. The dump plane is represented by a thick line on the vertical axis of sub-figures (a) and (d).}  
         \label{fig:Saddle_XY_Location_Compression_Expansion_SLPM750_MJ0_final}
\end{figure*} 

We see from Figs. \ref{fig:Chemi_Instability_wochemi_grayscale_SLPM750_MJ0_without_saddle_circle_labels}a-d that the saddle point $S_{1}$ emerges near the tip of the dump plane and remains in its close vicinity when $p'>0$. When $p'<0$ (Figs. \ref{fig:Chemi_Instability_wochemi_grayscale_SLPM750_MJ0_without_saddle_circle_labels}e-g), the saddle point subsequently advects upwards quickly and hits the top wall of the combustor along the shear layer. A new saddle point $S_{n}$ emerges near the dump plane at the end of an acoustic cycle and the whole process is then periodically repeated for every cycle of acoustic oscillation. 

We track the saddle points in the dump plane shear layer over 100 cycles of oscillations and obtain their trajectories during thermoacoustic instability. Figure \ref{fig:Saddle_XY_Location_Compression_Expansion_SLPM750_MJ0_final}a shows the location of saddle points at various instants during those 100 cycles. The saddle points are color-coded to indicate whether the acoustic pressure fluctuations are in compression phase (orange) or in expansion phase (blue). The overall spread of saddle points shows that they stay near the dump plane during the compression phase and advect downstream during the expansion phase. Distribution of the locations of saddle points in the vertical and horizontal directions are presented using probability density functions shown in Figs. \ref{fig:Saddle_XY_Location_Compression_Expansion_SLPM750_MJ0_final}b \& \ref{fig:Saddle_XY_Location_Compression_Expansion_SLPM750_MJ0_final}c respectively. The most probable location of saddle points (with a probability threshold $p(s)$= 0.05) during the compression phase is bound by light grey dashed lines in both Figs. \ref{fig:Saddle_XY_Location_Compression_Expansion_SLPM750_MJ0_final}b \& \ref{fig:Saddle_XY_Location_Compression_Expansion_SLPM750_MJ0_final}c. These bounded lines form the red box in Fig \ref{fig:Saddle_XY_Location_Compression_Expansion_SLPM750_MJ0_final}a. During the compression phase of thermoacoustic instability, most of the saddle points (66.1\%) lie inside the red box. After 40 acoustic cycles, the probability density function does not change, thereby not affecting the size of the red box. We therefore consider the red box as a region of importance to implement passive control.

The individual trajectories of these saddle points are shown in figure \ref{fig:Saddle_XY_Location_Compression_Expansion_SLPM750_MJ0_final}d. The trajectories of saddle points start at the dump plane and advect towards the top wall of the combustor in one cycle of oscillation. We obtain the trajectories of the saddle points along the shear layer for 100 acoustic cycles. These trajectories represent the location of the shear layer. Then, the curvilinear median of those trajectories gives the typical or central value of the distribution across y dimension as a function of x-dimension. The median trajectory (red line) is overlaid over the individual trajectories of saddle points which shows the most probable location of the oscillating shear layer. This shear layer creates a confined zone separating the outer recirculation zone with its trapped vortex from the core flow. The arclength of the median line of the saddle point trajectory (34.7 mm) is roughly equal to the distance between the dump plane and bluff-body (35 mm) which roughly corresponds to the distance traveled back and forth by the shear layer in the dump plane.

A positive Rayleigh Index is a necessary but not sufficient condition for thermoacoustic instability to occur in a system\citep{rayleigh, Dowling, Schuermans}. Rayleigh Index (R.I.) can be written as,
\begin{equation}
\label{eq:1}
\mathcal{R}=\dfrac{1}{T} \int_T p^\prime (x, t) \dot{q}^\prime (x, t) \,dt
\end{equation} 
\vspace{-25pt}

\begin{figure*}
\centering
\includegraphics[scale=0.42]{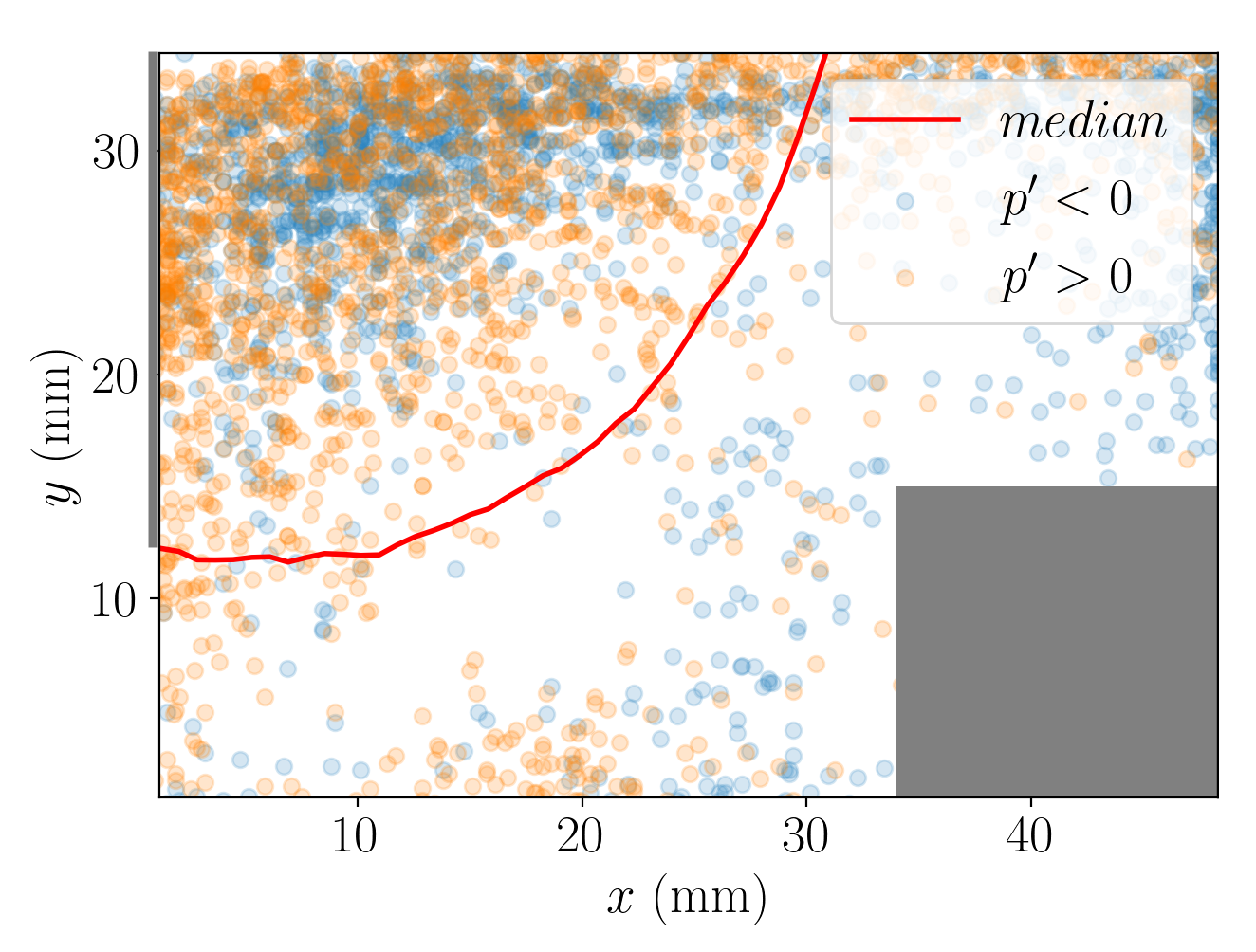}
\caption{The location of maximum heat release rate fluctuations in the compression phase ($p^\prime>0$) and the location of minimum heat release rate fluctuations in the expansion phase ($p^\prime<0$) during the state of thermoacoustic instability. The locations are concentrated in the outer recirculation zone (ORZ) near the dump plane: 71.8\% when $p' > 0$, 68.1\% when $p' < 0$. Overall 65.2\% of the locations are inside the ORZ. The dump plane is represented by a thick line on the vertical axis.} 
\label{fig:Flame_Combine_Max_Min_RI_pIN_pDE_XY_Location_SLPM750_MJ0_wopdf_final}
\end{figure*}

\noindent In this equation, the product of $p^\prime \dot{q}^\prime $ represents a thermoacoustic source term. A positive Rayleigh Index is the result of either positive heat release rate fluctuations during compression phase ($p^\prime>0$) of pressure or negative heat release rate fluctuations during the expansion phase ($p^\prime<0$) of the pressure. 

To understand the regions where the thermoacoustic source term is dominant, we track the spatial locations of maxima and minima of heat release rate fluctuations obtained as high speed $CH^*$ chemiluminescence images. In figure \ref{fig:Flame_Combine_Max_Min_RI_pIN_pDE_XY_Location_SLPM750_MJ0_wopdf_final}, we plot a subset of these extrema wherein the maximum heat release rate fluctuations obtained during the compression phase of $p^\prime$ are represented using orange dots and the minimum heat release rate fluctuations obtained during the expansion phase of $p^\prime$ are represented using blue dots. The median saddle point trajectory is overlaid in red to mark the most probable location of the oscillating shear layer that bounds the ORZ. In both the compression and expansion phases of $p^\prime$ during thermoacoustic instability, a majority of the spatial locations as outlined earlier lie inside the ORZ (71.8\% and 68.1\% respectively). In other words, the spatial locations where the Rayleigh index is positive mostly lie inside the ORZ. This ORZ, as we have seen using LCS, contains a vortex trapped by the free shear layer emerging from the dump plane. The analysis of the vortex and the free shear layer also revealed that they are receptive to pressure fluctuations inside the combustor, with periodic oscillations back and forth in phase with the pressure oscillations. This periodic dynamics of the shear layer and the trapped vortex which carry the flame produce periodic fluctuations in heat release rate that further driving the acoustic pressure fluctuations inside the combustor, thereby closing the feedback loop for thermoacoustic instability. 

\begin{figure*}
    \centering
    \includegraphics[scale=0.9]{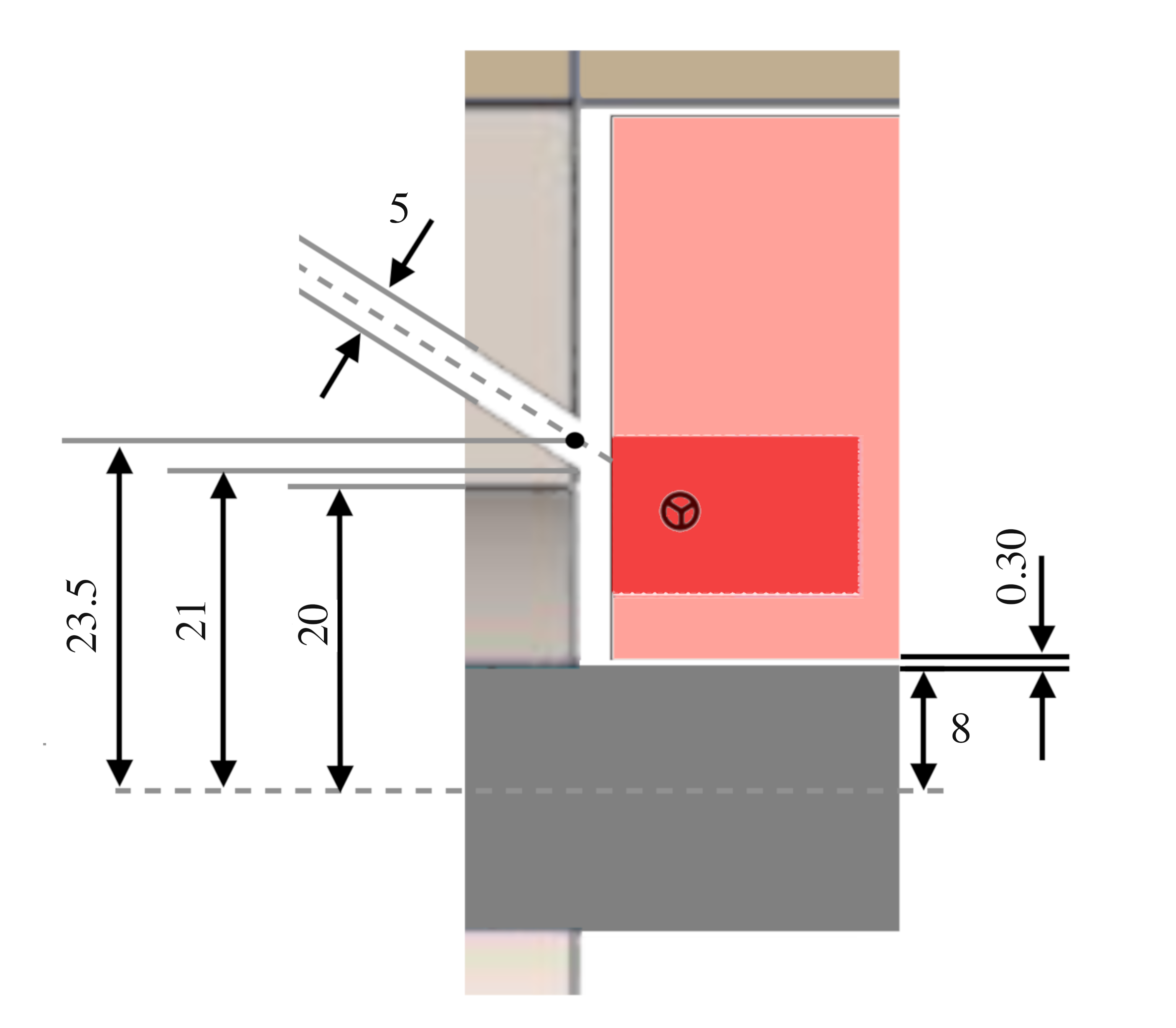}
    \caption{Schematic figure showing the dimensions of passive control port with respect to the dump plane. Red boxed region is the optimal location identified though a statistical analysis of saddle points. The most probable location of the saddle points is marked using a black tri-down.}
    \label{fig:Passive_control_pmt_pq_SLPM750_MJ0_MJ60_combined_mfc_time_FFTPQ_a_outinset_pdf_crop}
    \end{figure*}
    
\begin{figure*}
    \centering
    \includegraphics[scale=1]{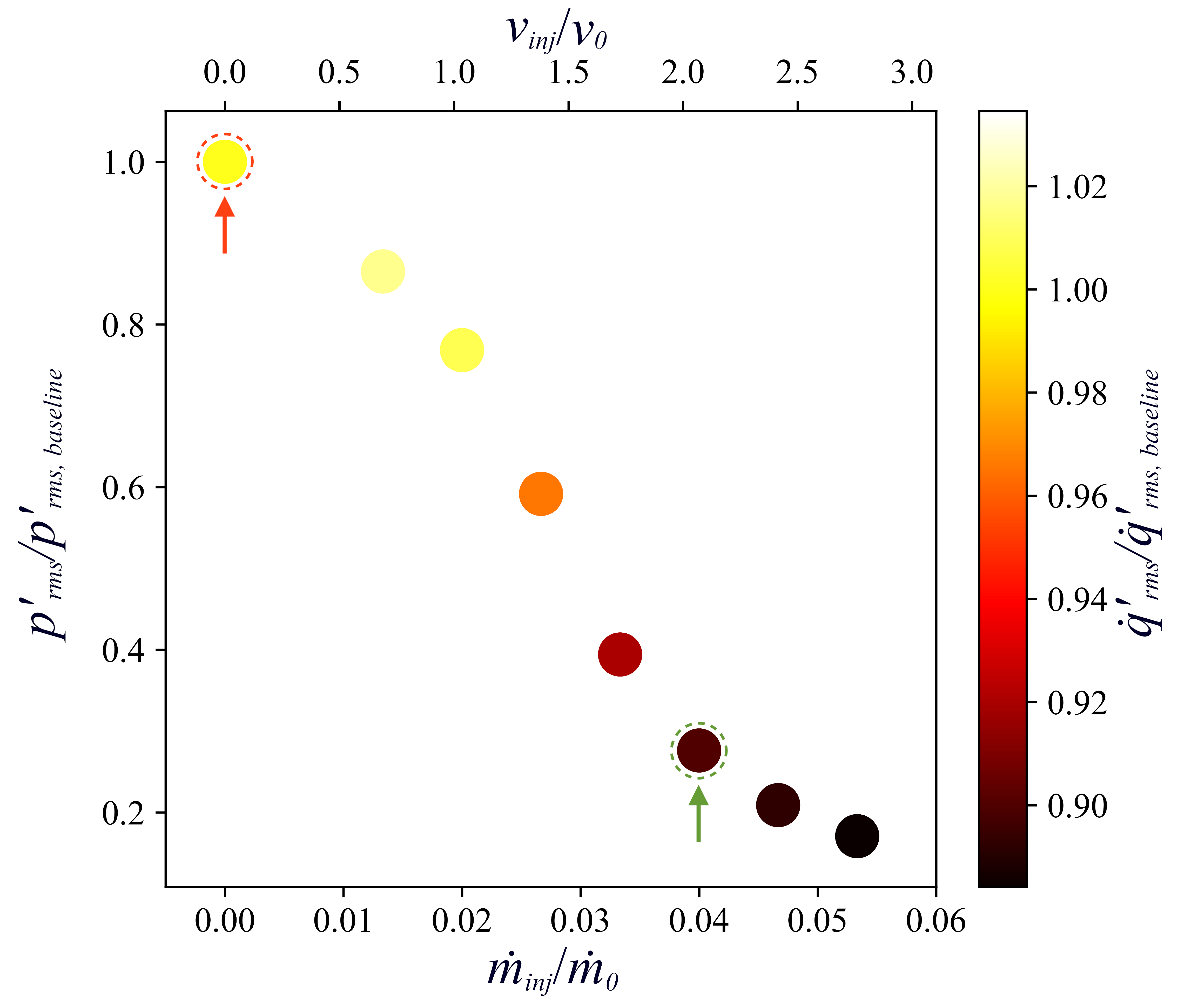}
    \caption{Variation of the ratio of root mean square (r.m.s.) value of pressure fluctuations with secondary micro-jet injection ($p'_{rms}$) to r.m.s. value of pressure fluctuations without secondary injection (baseline case) $p'_{rms, baseline}$ plotted against the ratio of average velocities ($v_{inj}/v_{0}$) and average mass flow rate ratio ($\dot{m}_{inj}/\dot{m}_{0}$). Color codes indicate the ratio of r.m.s. of global heat heat release rate ($\dot{q}'_{rms}$) to the baseline case. Baseline case is of thermoacoustic instability is indicated using a red arrow. The green arrow shows the suppressed state where further spatio-temporal analysis is performed.}
    \label{fig:Passive_control_pmt_pq_SLPM750_MJ0_MJ60_combined_mfc_time_FFTPQ_a}
    \end{figure*}    
\begin{figure*}
    \centering
    \includegraphics[scale=0.8]{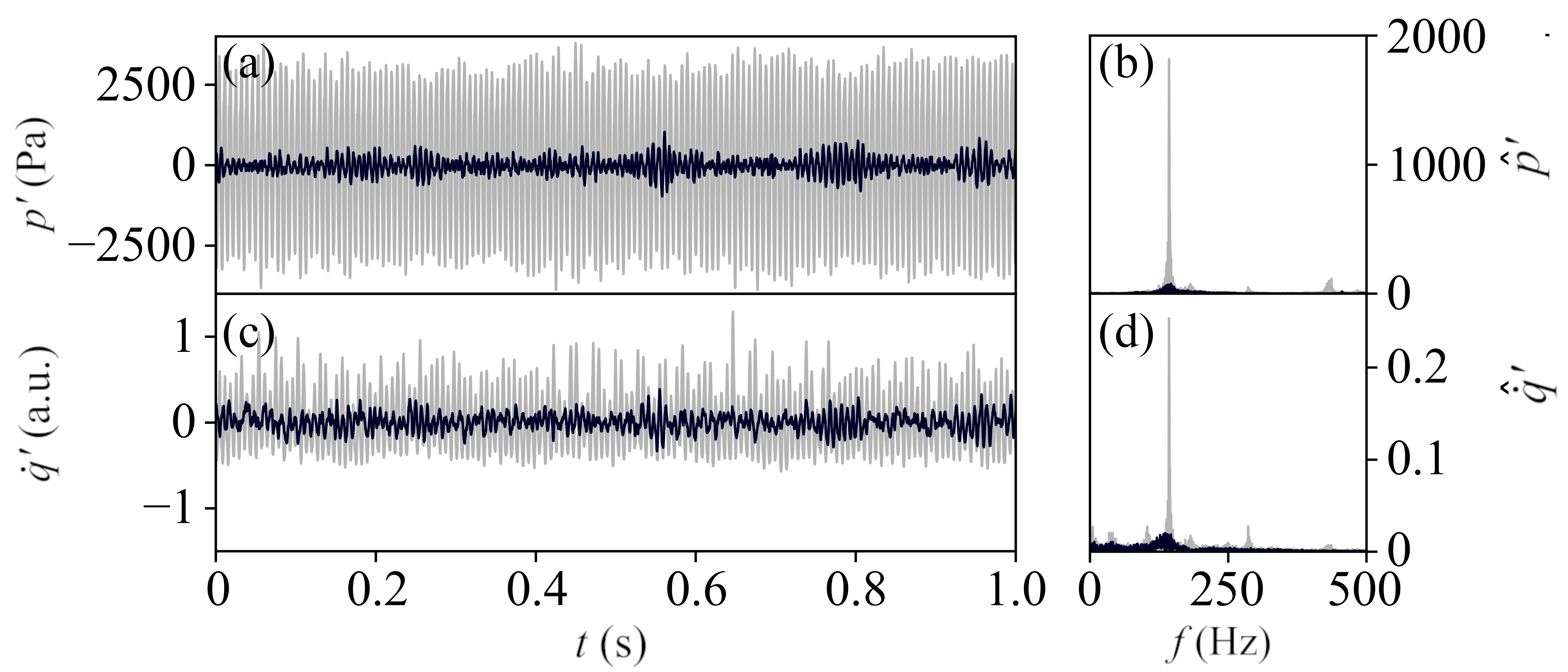}
    \caption{Time variation of (a) pressure and (c) global heat release rate fluctuations during the state of instability (gray) and suppressed state (black). Fast Fourier transform of the corresponding signals are shown in (b) and (d), respectively.}
    \label{fig:Passive_control_pmt_pq_SLPM750_MJ0_MJ60_combined_mfc_time_FFTPQ b labels}
    \end{figure*}

\subsection{Flow dynamics during the state of suppression of thermoacoustic instability}
\label{Flow dynamics in the suppressed state of passive control}
 \begin{table*}
    \caption{Parameter values at the operating conditions investigated.}
    \begin{center}
    \label{table:parameters}
    \begin{tabular}{c l l l l l}
    & & \\ 
    \hline
    Cases & $\phi$ & $v_{0}$ (m/s)  & $v_{inj}$ (m/s) &  $\dot{m}_{inj}$ (g/s) & $f_{a}$ (Hz) \\
    \hline
    Thermoacoustic instability (baseline) & 0.63 & 12.3 & 0 & 0 & 143.1 \\
    Control & 0.58 & 12.3 & 25.4 & 0.61 & - \\
    \hline
    \end{tabular}
    \end{center}
    \end{table*}

Experiments were conducted by providing a micro-jet port in the sidewall of the dump plane for passive control of thermoacoustic instability. A schematic diagram of the micro-jet port provision is shown in figure \ref{fig:Passive_control_pmt_pq_SLPM750_MJ0_MJ60_combined_mfc_time_FFTPQ_a_outinset_pdf_crop}. The most probable location of the saddle points identified in figure \ref{fig:Saddle_XY_Location_Compression_Expansion_SLPM750_MJ0_final} using the joint probability density function are marked in black tri-down (inside the red boxed region in figure \ref{fig:Passive_control_pmt_pq_SLPM750_MJ0_MJ60_combined_mfc_time_FFTPQ_a_outinset_pdf_crop}). The ports are designed at an angle of 60$^\circ$ (considering the fabrication difficulties caused by mechanical components) in both the top and bottom surfaces of the sidewall that focuses the micro-jet onto the most probable location. A steady micro-jet of air is continuously injected from ports of 5 mm diameter at the dump plane symmetrically from the top and bottom, focusing on the optimal location to disrupt the trajectory of the saddle points.

\begin{figure*}
    \centering
    \includegraphics[scale=0.24]{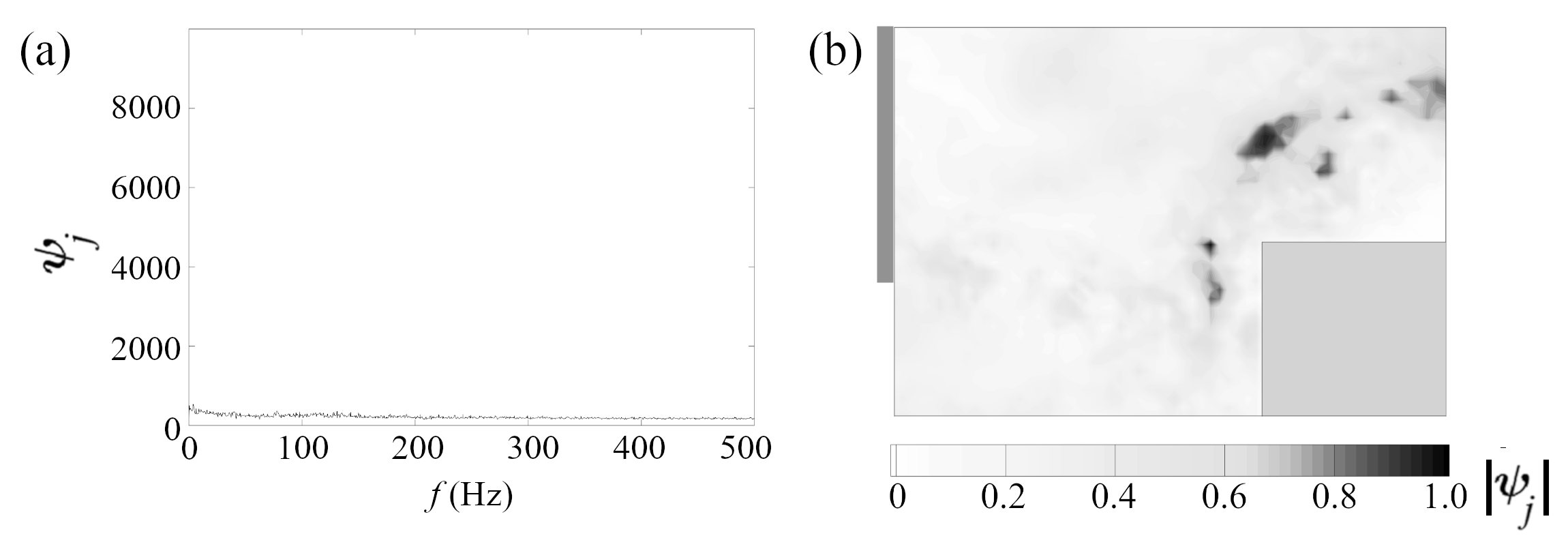}
    \caption{(a) Variation of the norm of dynamic modes with frequency. No sharp peaks are observed. (b) Normalized dynamic mode at the instability frequency $f_{a}$. There are no dominant structures near the dump plane as observed during the state of thermoacoustic instability. The dump plane is represented by a thick line on the vertical axis.}
    \label{fig:DMNorm_Freq_DM_MJ60}
\end{figure*}    

Passive control experiments were conducted during the state of thermoacoustic instability ($\phi$ = 0.63, $v_{0}$ = 12.3 m/s) for various mass flow rates of secondary micro-jet of air ranging from 0 to 0.82 g/s. The variation of ratio of root mean square (r.m.s.) values of the pressure signal to the baseline case of no secondary injection ($p'_{rms}/ p'_{rms, baseline}$), with the ratio of average injection velocities ($v_{inj}/v_{0}$) (or the average mass flow rate ratios ($\dot{m}_{inj}/\dot{m}_{0}$)) of the injector is shown in figure \ref{fig:Passive_control_pmt_pq_SLPM750_MJ0_MJ60_combined_mfc_time_FFTPQ_a} using circular markers. The ratio of the corresponding r.m.s. values of global heat release rate fluctuations compared to the baseline case is shown as the colored contour inside the circular markers. The baseline case of thermoacoustic instability is also marked using a red arrow in figure \ref{fig:Passive_control_pmt_pq_SLPM750_MJ0_MJ60_combined_mfc_time_FFTPQ_a}. Increasing the mass flow rate of the secondary micro-jet, there is a drastic decrease in the r.m.s of pressure fluctuations ($p'_{rms}$) comparable to the values observed during stable operation of the combustor without control action (Figure \ref{fig:Passive_control_pmt_pq_SLPM750_MJ0_MJ60_combined_mfc_time_FFTPQ_a}). However, the flame intensity slightly increases with an initial increase in the secondary mass flow rate of air. Upon further increasing the secondary mass flow rate of air, the intensity of the flame decreases. We select a passive control case (green arrow in figure \ref{fig:Passive_control_pmt_pq_SLPM750_MJ0_MJ60_combined_mfc_time_FFTPQ_a}) where the r.m.s. of pressure oscillation is comparatively lower for further analysis. Parameters for the baseline case and the selected case of passive control are summarized in table \ref{table:parameters}. We also injected a steady secondary micro-jet of air onto other locations both upstream and downstream of the bluff body where the suppression of sound is not observed. Figure 11 of Krishnan $et. al$ \cite{Abin2} show the level of suppression when injecting the secondary jet on the other locations.  

The time series of pressure and global heat release rate fluctuations in the absence of passive control (grey lines) and in the presence of passive control (black) are shown in figure \ref{fig:Passive_control_pmt_pq_SLPM750_MJ0_MJ60_combined_mfc_time_FFTPQ b labels}a \& c respectively.  We see high amplitude periodic oscillations that are manifested as tonal sound during thermoacoustic instability (indicated in grey). The amplitude of fluctuations is reduced drastically when control is implemented (indicated in black). Fast Fourier transform (FFT - grey lines) of pressure ($p^\prime$) and heat release rate fluctuations ($\dot{q}^\prime$) shows a sharp peak at acoustic frequency $f_{a}$ = 143.1 Hz during thermoacoustic instability (see figure \ref{fig:Passive_control_pmt_pq_SLPM750_MJ0_MJ60_combined_mfc_time_FFTPQ b labels}b \& d). The FFTs of $p^\prime$ (black in figure \ref{fig:Passive_control_pmt_pq_SLPM750_MJ0_MJ60_combined_mfc_time_FFTPQ b labels}b) and $\dot{q}^\prime$ after passive control (black in figure \ref{fig:Passive_control_pmt_pq_SLPM750_MJ0_MJ60_combined_mfc_time_FFTPQ b labels}d) show that the acoustic frequency ($f_{a}$) has been suppressed completely. Hardalupas and Orain\citep{hardalupas} related the chemiluminscence with heat release rate fluctations. We nondimensionalize as it does not change the trend or frequency of the signal. We observe a reduced pressure amplitude of $p'_{rms}=238.3 \pm 0.97$ Pa with control as compared to the pressure amplitude observed during thermoacoustic instability ($p'_{rms}=2198.3 \pm 8.97$ Pa). This drastic 90\% reduction in the amplitude is achieved with a secondary jet having a mass flow rate of just 4\% of the primary flow. At this lower global equivalence ratio of 0.58 (see table \ref{table:parameters}), the combustor was originally unstable without secondary micro-jet injection. 

The velocity field acquired for the suppressed state of thermoacoustic instability ($\phi$ = 0.58, $v_{0}$ = 12.3 m/s, $v_{inj}$ = 25.4 m/s, $T_0$ = 30 ms) is decomposed using DMD. The variation of the norm of the dynamic modes ($\psi_{j}$) with frequency ($f$(Hz)) shown in figure \ref{fig:DMNorm_Freq_DM_MJ60}a reveal that there are no longer any dominant frequencies in the flow dynamics as was observed during thermoacoustic instability. The normalized dynamic mode at acoustic frequency ($f_{a}$) is plotted in figure \ref{fig:DMNorm_Freq_DM_MJ60}b. The dominant structure observed in the dump plane during thermoacoustic instability has disappeared completely upon injection of secondary micro-jet of air.

\begin{figure*}
\centering
\includegraphics[scale=2.6]{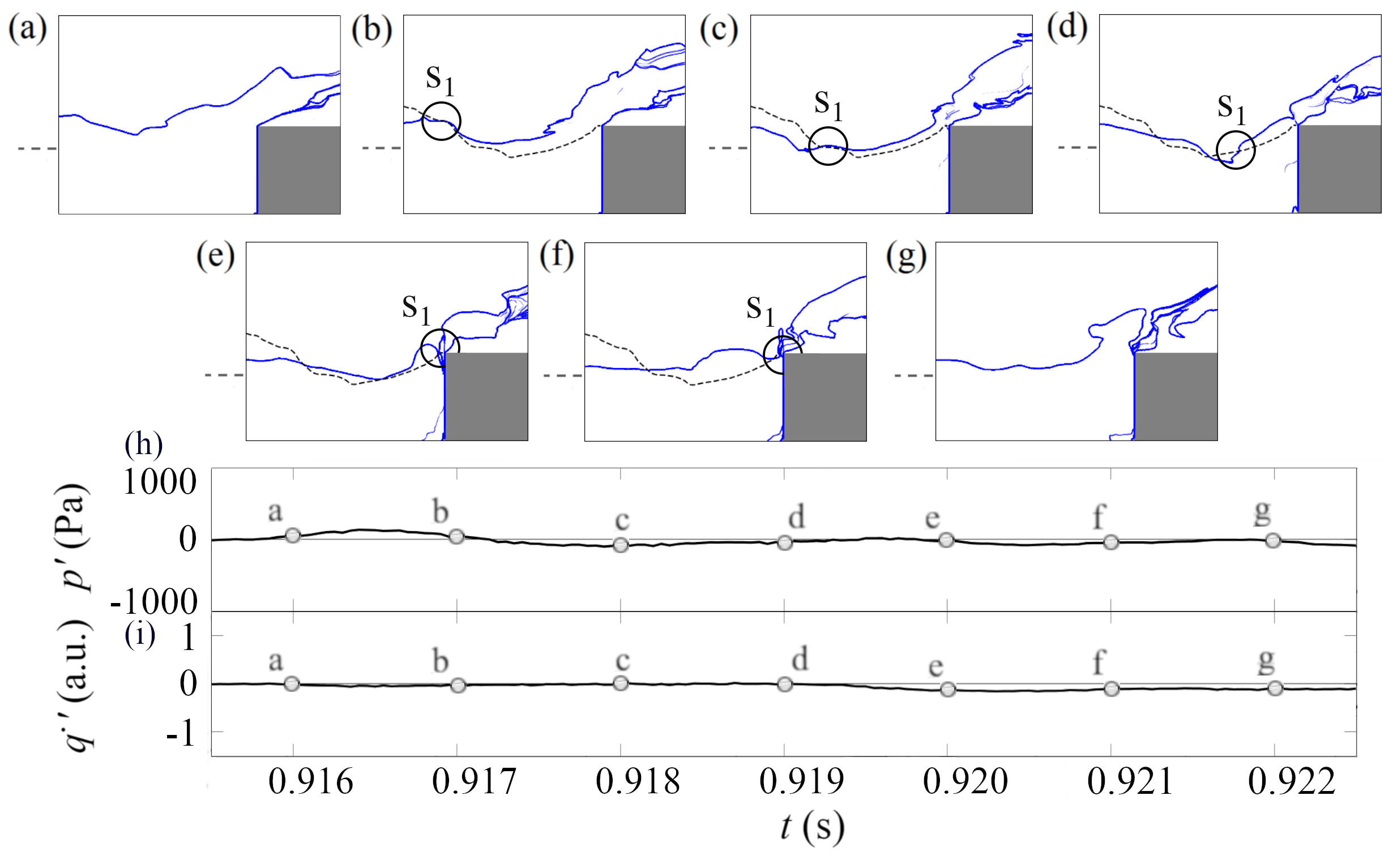}
\caption{(a-g) Attracting LCS or coherent structures (blue) after performing passive control by symmetrically injecting a secondary micro-jet of air ($\phi$ = 0.58, $v_{0}$ = 12.3 m/s, $v_{inj}$ = 25.5 m/s, $T_0$= 30 ms). The few snapshots are shown here during the suppressed state of thermoacoustic instability. The saddle points are marked using black circles. The trajectory of a sample saddle point $S_{1}$ is shown using a dashed black line. Unsteady (h) pressure and (i) heat release rate fluctuation signals corresponding to the time instants of the snapshots. The dump plane is represented by the horizontal dashed line on the vertical axis of each snapshot. The flow is from left to right.}
\label{fig:Chemi_Instability_wochemi_grayscale_SLPM750_MJ60_without_saddle_circle_labels}
\end{figure*}    

\begin{figure*}
    \centering
    \includegraphics[scale=0.42]{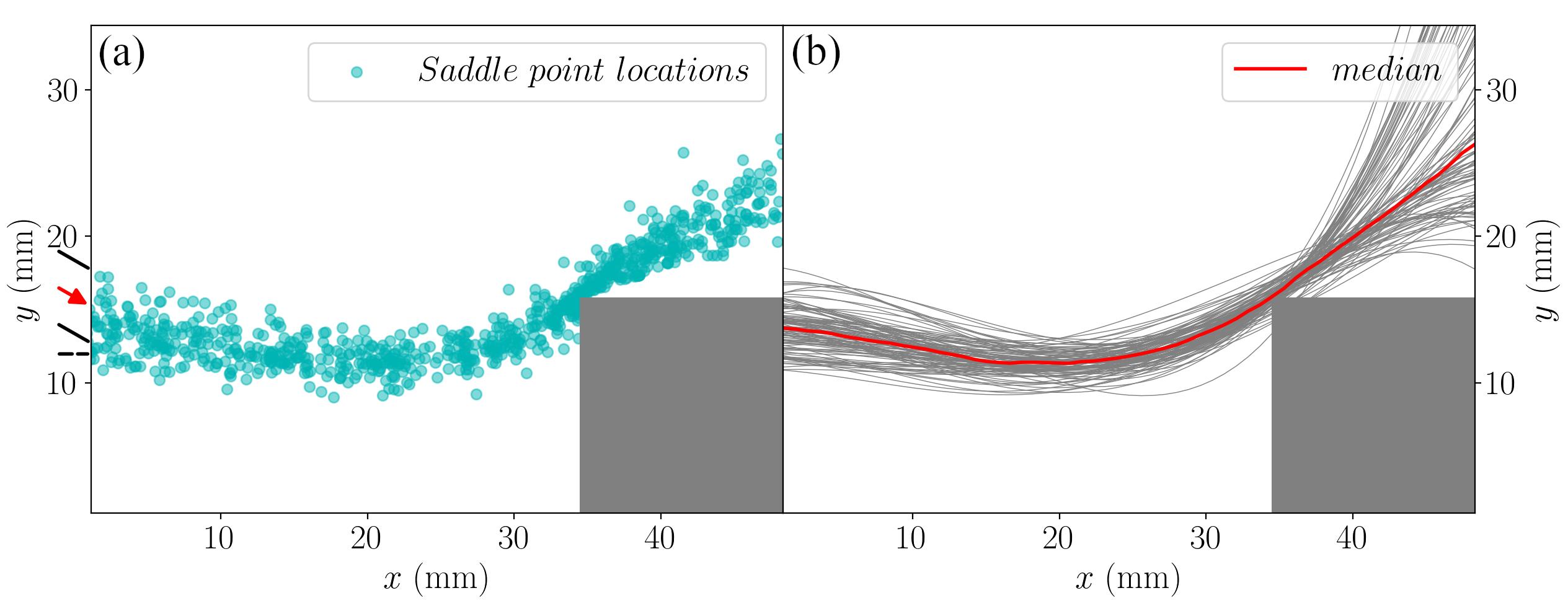}
    \caption{(a) Location of saddle points over 100 acoustic cycles after injecting the secondary micro-jet of air ($\phi$ = 0.58, $v_{0}$ = 12.3 m/s, $v_{inj}$ = 25.5 m/s, $T_0$= 30 ms), and (b) individual trajectories (N=100) of saddle points along the shear layer. Median of the individual trajectories is overlaid in red. The saddle point trajectories advect almost parallel to the horizontal axis of the combustor in the presence of control action.}
    \label{fig:Combined_Saddle_trajectories_Mean_Std_SLPM750_MJ60_black}
\end{figure*}   
Repelling LCS overlaid over attracting LCS during the suppressed state ($\dot{m}_{inj}/\dot{m}_{0}$ = 0.04) over an acoustic cycle of $T_{a} = 1/f_{a}$ are presented in figure \ref{fig:Chemi_Instability_wochemi_grayscale_SLPM750_MJ60_without_saddle_circle_labels}a-g. The corresponding time signals of pressure and heat release rate fluctuations with the snapshot instances marked are also presented in Figs. \ref{fig:Chemi_Instability_wochemi_grayscale_SLPM750_MJ60_without_saddle_circle_labels}h,i. In preparing these figures, time instances are taken to be the same as that observed without control. In every snapshot, the saddle point ($S_{i}$) when present is marked using a black circle.

We observe that attracting LCS display considerably less complexity in the suppressed state. For instance, the vortices which were originally trapped in the outer recirculation zone are no longer visible as part of the attracting LCS. Specifically, the attracting LCS representing the free shear layer emerging from the dump plane is pushed downwards to make it almost parallel to the combustor axis by the secondary micro-jet. Since there is no concentration pf vorticity in the ORZ, the shear layer motion is no longer modulated by the back-and-forth motion of the trapped vortices as was happening during thermoacoustic instability. Consequently, the oscillations in the shear layer are suppressed and we have control of thermoacoustic instability.

We see that the saddle points (for instance, $S_{1}$) after emerging from the dump plane are forced to advect almost parallel to the horizontal axis of the combustor by the secondary micro-jet (Figs. \ref{fig:Chemi_Instability_wochemi_grayscale_SLPM750_MJ60_without_saddle_circle_labels}b - \ref{fig:Chemi_Instability_wochemi_grayscale_SLPM750_MJ60_without_saddle_circle_labels}f). This is not surprising as it is the saddle point dynamics that dictate shear layer dynamics.  When the saddle points are forced to alter their trajectories by secondary injection, the shear layer gets consequently pushed downwards. 

As was performed during thermoacoustic instability, we track the location of the saddle points in each snapshot and compute individual saddle point trajectories in the suppressed state ($\phi$ = 0.58, $v_{0}$ = 12.3 m/s, $v_{inj}$ = 25.4 m/s) over 100 acoustic cycles. Figure \ref{fig:Combined_Saddle_trajectories_Mean_Std_SLPM750_MJ60_black}a shows the spread of saddle points emerging from the dump plane and advecting downstream. Trajectories of individual saddle points are shown in figure \ref{fig:Combined_Saddle_trajectories_Mean_Std_SLPM750_MJ60_black}b. In figure \ref{fig:Combined_Saddle_trajectories_Mean_Std_SLPM750_MJ60_black}b, the trajectories of individual saddle points are depicted in red, and a median line is overlaid to indicate the most likely position of the shear layer during the suppressed state. We find that there is considerably less deviation in the individual trajectories from the median trajectory compared to what was observed during the state of thermoacoustic instability.  Unlike the state of thermoacoustic instability, the saddle points are no longer able to advect upwards and hit the top wall of the combustor, making the shear layer lie almost horizontal to the combustor axis. 

We have previously shown that there are two shear layers, one emerging from the dump plane and another emerging from the leading edge of the bluff-body that synchronously come closer during the compression phase of pressure and move apart during the expansion phase of pressure \citep{prem_pof}. This synchronous motion forces trajectories of the fluid parcels to converge ($p^\prime>0$) and diverge ($p^\prime<0$) in the region between the shear layers (figure \ref{fig:Chemi_Instability_wochemi_grayscale_SLPM750_MJ0_without_saddle_circle_labels}). Based on the DMD and LCS analysis performed, we can conclude that the periodicity in the shear layer emanating from the dump plane and the trapped vortex forces periodicity onto the shear layer emerging from the leading edge of the bluff-body producing this periodic convergence and divergence of trajectories.

\section{CONCLUSIONS}
\label{CONCLUSIONS}
In the current study, we propose a patented framework to identify the optimal location for passive control action to mitigate sound production due to thermoacoustic instability. 
We observe\textemdash through experiments and statistical analysis in a bluff-body stabilized combustor\textemdash that during thermoacoustic instability, the outer recirculation zone (ORZ) is bounded by the free shear layer that originates from the dump plane and stretches to the top wall of the combustor. This free shear layer is obtained by evolving trajectories containing the saddle points. The vortex and bounding shear layer are receptive to pressure fluctuations inside the combustor and periodically move back and forth in phase with the acoustic pressure oscillations. Since the shear layer and the trapped vortex carry the flame, their periodic oscillations lead to periodic heat release rate fluctuations that drive the acoustic field. Even when there are intense, high-amplitude heat release rate and pressure oscillations during thermoacoustic instability, the vortex in the ORZ remains trapped. 

The optimal location for passive control action is identified as the region in the free shear layer where the saddle points spend a majority of their time during an acoustic cycle on an average.
We inject a steady micro-jet of air onto this optimal location and achieve a suppression of sound of almost $90\%$ with a micro-jet mass flow rate of just 4\% of the primary flow. On inspecting the shear layer dynamics at the optimal location after control, we find that the trajectory containing saddle points is diverted by the micro-jet, making it advect almost parallel to the combustor axis. We observed no suppression of sound upon implementing passive control action on other locations both upstream and downstream of the bluff body. 

The techniques presented in this paper serve as a proof-of-concept towards identifying optimal locations for passive control so that large amplitude oscillations produced during the state of thermoacoustic instability can be suppressed. Through careful consideration, they can also be used to glean into the mechanisms of thermoacoustic instability at operating conditions well before thermoacoustic instability sets in\citep{prem_pof, prem_asme}. However, successful implementation of such strategies requires time-resolved velocity data that are difficult to obtain experimentally for complex geometries, where one may alternately rely on numerical simulations.

\setcounter{secnumdepth}{0} 

\section*{Disclosure statement}
\label{Disclosure statement}
The authors report no conflict of interest.

\section*{Acknowledgments}
\label{Acknowledgments}
C.P. Premchand and Vineeth Nair would like to thank SERB (DST) for funding the study through CRG (Grant no.: CRG/2020/003406). The authors wish to thank Mr. S. Thilagaraj and Mr. S. Anand (Indian Institute of Technology Madras) for their support in performing the experiments. We also thank our colleagues from the IIT Bombay (Dr. Pranav Thakare) and IIT Madras (Dr. Amitesh Roy, Dr. Induja Pavithran) for their critical comments and suggestions. R.I. Sujith thanks SERB (DST) for Swarnajayanti (DST/SF/1(E C)/2006) and J C Bose fellowships (JCB/2018/000034/SSC).


\begin{thebibliography}{16}

\bibitem{prem_pof}Premchand, C.P., George, N., Raghunathan, M., Unni, V., Sujith, R.I. \& Nair, V. Lagrangian analysis of intermittent sound sources in the flow-field of a bluff-body stabilized combustor. {\em Physics of Fluids}. \textbf{31}, 025115 (2019)

\bibitem{rayleigh}Rayleigh, L. The explanation of certain acoustical phenomena. {\em Roy. Inst. Proc.}. \textbf{8} pp. 536-542 (1878)

\bibitem{Juniper}Juniper, M. \& Sujith, R.I. Sensitivity and nonlinearity of thermoacoustic oscillations. {\em Annual Review of Fluid Mechanics}. \textbf{50} pp. 661-689 (2018)

\bibitem{Sutton}Sutton, G. History of liquid-propellant rocket engines in Russia, formerly the Soviet Union. {\em Journal of Propulsion and Power}. \textbf{19}, 1008-1037 (2003)

\bibitem{Lieuwen2}Lieuwen, T. \& Yang, V. Combustion instabilities in gas turbine engines: operational experience, fundamental mechanisms, and modeling. (American Institute of Aeronautics,2005)

\bibitem{Biggs}Biggs, R. Rocketdyne–F-1 Saturn first stage engine. {\em Remembering The Giants: Apollo Rocket Propulsion Development}. pp. 15-26 (2009)

\bibitem{Nair5}Nair, V., Thampi, G. \& Sujith, R.I. Intermittency route to thermoacoustic instability in turbulent combustors. {\em Journal of Fluid Mechanics}. \textbf{756} pp. 470-487 (2014)

\bibitem{nair}Nair, V. \& Sujith, R.I. Multifractality in combustion noise: predicting an impending combustion instability. {\em Journal of Fluid Mechanics}. \textbf{747} pp. 635-655 (2014)

\bibitem{Candel}Candel, S., Durox, D., Ducruix, S., Birbaud, A., Noiray, N. \& Schuller, T. Flame dynamics and combustion noise: progress and challenges. {\em International Journal of Aeroacoustics}. \textbf{8}, 1-56 (2009)

\bibitem{Tandon}Tandon, S., \& Sujith, R.I. Multilayer network analysis to study complex inter-subsystem interactions in a turbulent thermoacoustic system. {\em Journal of Fluid Mechanics}, \textbf{966}, A9 (2023). doi:10.1017/jfm.2023.338

\bibitem{Sujith_PoF}Sujith, R.I. \& Unni, V. Complex system approach to investigate and mitigate thermoacoustic instability in turbulent combustors. {\em Physics of Fluids}. \textbf{32}, 061401 (2020)

\bibitem{Sujith_proceedings}Sujith, R.I. \& Unni, V. Dynamical systems and complex systems theory to study unsteady combustion. {\em Proceedings of the Combustion Institute}. \textbf{38}, 3445-3462 (2021)

\bibitem{gotoda}Gotoda, H., Shinoda, Y., Kobayashi, M., Okuno, Y. \& Tachibana, S. Detection and control of combustion instability based on the concept of dynamical system theory. {\em Physical Review E}. \textbf{89}, 022910 (2014)

\bibitem{Ebi}Ebi, D., Denisov, A., Bonciolini, G., Boujo, E. \& Noiray, N. Flame dynamics intermittency in the bistable region near a subcritical Hopf bifurcation. {\em Journal of Engineering for Gas Turbines and Power}. \textbf{140} (2018)

\bibitem{kheirkhah}Kheirkhah, S., Cirtwill, J., Saini, P., Venkatesan, K. \& Steinberg, A. Dynamics and mechanisms of pressure, heat release rate, and fuel spray coupling during intermittent thermoacoustic oscillations in a model aeronautical combustor at elevated pressure. {\em Combustion and Flame}. \textbf{185} pp. 319-334 (2017)

\bibitem{sampath_CaF}Sampath, R. \& Chakravarthy, S. Investigation of intermittent oscillations in a premixed dump combustor using time-resolved particle image velocimetry. {\em Combustion and Flame}. \textbf{172} pp. 309-325 (2016)

\bibitem{Zhao1}Zhao, D. \& Li, X. A review of acoustic dampers applied to combustion chambers in aerospace industry. {\em Progress in Aerospace Sciences}. \textbf{74} pp. 114-130 (2015)

\bibitem{Zhao2}Zhao, D., Lu, Z., Zhao, H., Li, X., Wang, B. \& Liu, P. A review of active control approaches in stabilizing combustion systems in aerospace industry. {\em Progress in Aerospace Sciences}. \textbf{97} pp. 35-60 (2018)

\bibitem{McManus}McManus, K., Poinsot, T. \& Candel, S. A review of active control of combustion instabilities. {\em Progress in Energy and Combustion Science}. \textbf{19}, 1-29 (1993)

\bibitem{Ghoniem}Ghoniem, A., Annaswamy, A., Park, S. \& Sobhani, Z. Stability and emissions control using air injection and H2 addition in premixed combustion. {\em Proceedings of the Combustion Institute}. \textbf{30}, 1765-1773 (2005)

\bibitem{Oztarlik}Oztarlik, G., Selle, L., Poinsot, T. \& Schuller, T. Suppression of instabilities of swirled premixed flames with minimal secondary hydrogen injection. {\em Combustion And Flame}. \textbf{214} pp. 266-276 (2020)

\bibitem{Schadow1}Schadow, K. \& Gutmark, E. Combustion instability related to vortex shedding in dump combustors and their passive control. {\em Progress in Energy and Combustion Science}. \textbf{18}, 117-132 (1992)

\bibitem{Altay_CaF}Altay, H., Hudgins, D., Speth, R., Annaswamy, A. \& Ghoniem, A. Mitigation of thermoacoustic instability utilizing steady air injection near the flame anchoring zone. {\em Combustion and Flame}. \textbf{157}, 686-700 (2010)

\bibitem{Tachibana}Tachibana, S., Zimmer, L., Kurosawa, Y. \& Suzuki, K. Active control of combustion oscillations in a lean premixed combustor by secondary fuel injection coupling with chemiluminescence imaging technique. {\em Proceedings of the Combustion Institute}. \textbf{31}, 3225-3233 (2007)

\bibitem{Unni_chaos}Unni, V., Krishnan, A., Manikandan, R., George, N., Sujith, R.I., Marwan, N. \& Kurths, J. On the emergence of critical regions at the onset of thermoacoustic instability in a turbulent combustor. {\em Chaos: An Interdisciplinary Journal of Nonlinear Science}. \textbf{28}, 063125 (2018)

\bibitem{Abin}Krishnan, A., Sujith, R.I., Marwan, N. \& Kurths, J. On the emergence of large clusters of acoustic power sources at the onset of thermoacoustic instability in a turbulent combustor. {\em Journal of Fluid Mechanics}. \textbf{874} pp. 455-482 (2019)

\bibitem{Abin1}Krishnan, A., Manikandan, R., Midhun, P., Reeja, K., Unni, V., Sujith, R.I., Marwan, N. \& Kurths, J. Mitigation of oscillatory instability in turbulent reactive flows: A novel approach using complex networks. {\em EPL (Europhysics Letters)}. \textbf{128}, 14003 (2019)

\bibitem{Abin2}Krishnan, A., Sujith, R.I., Marwan, N. \& Kurths, J. Suppression of thermoacoustic instability by targeting the hubs of the turbulent networks in a bluff body stabilized combustor. {\em Journal of Fluid Mechanics}. \textbf{916} A20. doi: 10.1017/jfm.2021.166. (2021)

\bibitem{Roy_CaF2}Roy, A., Premchand, C.P., Raghunathan, M., Krishnan, A., Nair, V. \& Sujith, R.I. Critical region in the spatiotemporal dynamics of a turbulent thermoacoustic system and smart passive control. {\em Combustion and Flame}. \textbf{226} pp. 274-284 (2021)

\bibitem{Haller}Haller, G. Lagrangian coherent structures. {\em Annu. Rev. Fluid Mech}. \textbf{47}, 137-162 (2015)

\bibitem{prem_patent}Premchand, C.P., Nair, V., Sujith, R.I., George, N., Raghunathan, M., \& Unni, V. System and method for optimizing passive control strategies of oscillatory instabilities in turbulent systems using finite-time Lyapunov exponents. U.S. Patent No. 11,378,488. Washington, DC: U.S. Patent and Trademark Office, issued July 5, 2022.

\bibitem{Sampath} Sampath, R., Mathur, M., and Chakravarthy, S. R., 
Lagrangian coherent structures during combustion instability in a premixed-flame backward-step combustor, {\em Phys. Rev. E.} \textbf{94}, 062209-p1 - 062209-p10 (2016).

\bibitem{prem_asme}Premchand, C.P., George, N., Raghunathan, M., Unni, V., Sujith, R.I. \& Nair, V. Lagrangian analysis of flame dynamics in the flow field of a bluff body-stabilized combustor. {\em Journal of Engineering for Gas Turbines and Power}. \textbf{142}, 011015 (2020)

\bibitem{Farazmand}Farazmand, M. \& Haller, G. Computing Lagrangian coherent structures from their variational theory. {\em Chaos: An Interdisciplinary Journal of Nonlinear Science}. \textbf{22}, 013128 (2012)

\bibitem{hardalupas}Hardalupas, Y. \& Orain, M. Local measurements of the time-dependent heat release rate and equivalence ratio using chemiluminescent emission from a flame. {\em Combustion and Flame}. \textbf{139}, 188-207 (2004)

\bibitem{nitin}George, N., Unni, V., Raghunathan, M. \& Sujith, R.I. Pattern formation during transition from combustion noise to thermoacoustic instability via intermittency. {\em Journal of Fluid Mechanics}. \textbf{849} pp. 615-644 (2018)

\bibitem{poinsot}Poinsot, T., Trouve, A., Veynante, D., Candel, S. \& Esposito, E. Vortex-driven acoustically coupled combustion instabilities. {\em Journal of Fluid Mechanics}. \textbf{177} pp. 265-292 (1987)

\bibitem{chakravarthy}Chakravarthy, S., Sivakumar, R. \& Shreenivasan, O. Vortex-acoustic lock-on in bluff-body and backward-facing step combustors. {\em Sadhana}. \textbf{32}, 145-154 (2007)


\bibitem{Schmid}Schmid, P.J. Dynamic mode decomposition of numerical and experimental data. {\em Journal of Fluid Mechanics}. \textbf{656} pp. 5-28 (2010)

\bibitem{Schmid1}Schmid, P.J., Li, L., Juniper, M.P. \& Pust, O. Applications of the dynamic mode decomposition. {\em Theoretical and Computational Fluid Dynamics}. \textbf{25}, 249-259 (2011)

\bibitem{Dowling}Dowling, A. P., \& Morgans, A. S. Feedback control of combustion oscillations. {\em Annu. Rev. Fluid Mech.}, \textbf{37}, 151-182 (2005)

\bibitem{Schuermans} Schuermans, B., Moeck, J., Blondé, A., Dharmaputra, B., \& Noiray, N. The Rayleigh integral is always positive in steadily operated combustors. {\em Proceedings of the Combustion Institute}, \textbf{39}, 4661-4669 (2023)

\end{thebibliography}
\end{document}